\documentclass[fleqn]{2020SCGE}
\usepackage{color}
\usepackage{float}
\usepackage{natbib}
\usepackage{gensymb}
\usepackage{graphicx}
\usepackage{tablefootnote}
\usepackage[toc]{multitoc}
\usepackage{rotating}
\usepackage{lscape}
\usepackage{pdflscape}
\usepackage{setspace}
\usepackage{booktabs} 
\usepackage{siunitx} 
\usepackage{xspace}

\usepackage{url}
\usepackage{hyperref}

\newcommand{\hi}{H{\,\sc i}\xspace}
\newcommand{\HI}{H{\,\sc i}\xspace}
\newcommand{\hj}{\text{\hi}}

\newcommand{\mjyb}{{\rm mJy~beam}^{-1}}

\newcommand{\mjybkms}{{\rm mJy~km~s}^{-1}}

\newcommand{\dotmin}{\rlap.{'}}
\newcommand{\dotdeg}{\rlap.{^\circ}}

\newcommand{\msun}{M_{\odot}}

\newcommand{\arcmin}{$^{\prime}$}
\newcommand{\arcsec}{$^{\prime\prime}$}
\newcommand{\kms}{\rm km\,s^{-1}}

\begin{document}

\ensubject{subject}

\ArticleType{Article}
\Year{2026}
\Month{January}
\Vol{8}
\No{8}
\DOI{8}
\ArtNo{8}
\ReceiveDate{May 1, 2026}
\AcceptDate{June 30, 2026}
\OnlineDate{July 15, 2026}

\title{The FAST All Sky \hi Survey DR2: \\
the FASHI Catalog and the \hi Mass Function}{FASHI}
\author[1,2,\footnote{Corresponding authors (Chuan-Peng Zhang: \url{cpzhang@nao.cas.cn}; Ming Zhu: \url{mz@nao.cas.cn}; Peng Jiang: \url{pjiang@nao.cas.cn}; Hong Guo: \url{guohong@shao.ac.cn})}]{Chuan-Peng Zhang}{}%
\author[1,2,$^\ast$]{Ming Zhu}{}%
\author[1,2,$^\ast$]{Peng Jiang}{}%
\author[3,$^\ast$]{Hong Guo}{}%
\author[1,2]{Jin-Long Xu}{}%
\author[1,2]{\\Xiao-Lan Liu}{}%
\author[1,2]{Nai-Ping Yu}{}%
\author[4]{Cheng Cheng}{}%
\author[5]{Jing Wang}{}%
\author[1]{Jie Wang}{}%
\author[]{FAST Collaboration}{}%
\AuthorMark{Zhang C.-P.}

\AuthorCitation{Zhang C.-P., et al}

\address[1]{State Key Laboratory of Radio Astronomy and Technology, National Astronomical Observatories, Chinese Academy of Sciences, Beijing 100101, China}
\address[2]{Guizhou Radio Astronomical Observatory, Guizhou University, Guiyang 550000, China}
\address[3]{Shanghai Astronomical Observatory, Chinese Academy of Sciences, Nandan Road 80, Shanghai 200030, China}
\address[4]{Chinese Academy of Sciences South America Center for Astronomy, National Astronomical Observatories, CAS, Beijing 100101, China}
\address[5]{Kavli Institute for Astronomy and Astrophysics, Peking University, Beijing 100871, China}

\abstract{
The FAST All Sky \hi Survey (FASHI) conducted with the Five-hundred-meter Aperture Spherical radio Telescope (FAST) has mapped  $\sim$19\,500\,deg$^2$ of the sky north of $\rm DEC = -14^\circ$, detecting 156\,411 extragalactic \hi sources at $z < 0.09$ with a median sensitivity of $0.57$\,mJy\,beam$^{-1}$ at a velocity resolution of $6.4$\,km\,s$^{-1}$. The survey achieves unprecedented depth and area coverage, significantly improving upon previous single-dish surveys. Through a detailed completeness analysis that accounts for the survey's non-uniform sensitivity and line-width dependence, we construct a robust \hi mass function (HIMF) using a completeness‑corrected sample of over 109\,000 sources. The HIMF is robustly constrained down to $M_{\text{HI}} \sim 10^{6.2} M_{\odot}$. When systematic uncertainties are included, the HIMF is well described by a single-Schechter function with a characteristic mass $\log(M_{s}/h_{70}^{-2}\msun) = 9.89 \pm 0.02$, low-mass end slope $\alpha=-1.31\pm0.02$ and amplitude $\phi_s=(6.38\pm0.49)\times10^{-3}h_{70}^3{\rm Mpc}^{-3}{\rm dex}^{-1}$. The derived cosmic \hi density is $\Omega_\hj = (4.71 \pm 0.03_{\rm stat}\pm0.40_{\rm sys}) \times 10^{-4} h_{70}^{-1}$. FASHI provides the most extensive and sensitive \hi catalog to date, establishing an important benchmark for studies of gas accretion, galaxy evolution, and large-scale structure in the local universe.
}


\maketitle

\begin{multicols}{2}

\section{Introduction}
\label{sect:intro}

Extragalactic neutral hydrogen (\hi), traced by the 21\,cm line, is a fundamental component of galaxies and a key regulator of gas accretion, star formation, and feedback across cosmic time. As the primary reservoir of cold gas available for star formation, the abundance and distribution of \hi provide critical constraints on galaxy evolution models and the baryon cycle between galaxies and the intergalactic medium. Blind \hi surveys therefore play a central role in establishing a statistical census of gas-rich galaxies and their environments \citep[e.g.,][]{Cheng2020,Cheng2025,Xu2022,Springob2005,Koribalski2004,Meyer2004,Wong2006,Giovanelli2015,Haynes2011,Haynes2018}.

\begin{figure*}
\centering
\includegraphics[width=0.99\textwidth, angle=0]{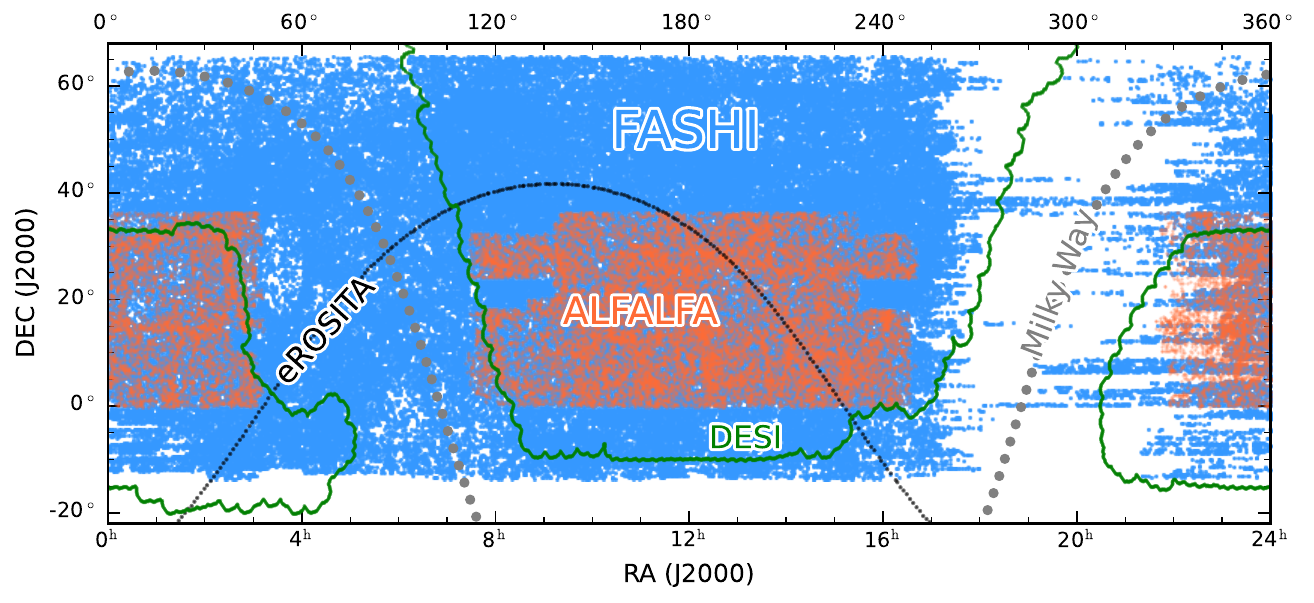}
\caption{Equatorial sky distribution of the 156\,411 FASHI \hi sources (blue points; see Table\,\ref{tab:exgalcat}), showing the patchy coverage resulting from observational constraints. For comparison, ALFALFA $\alpha$100 sources \citep{Haynes2018} are overlaid as orange points. The footprints of the DESI \citep{desi2025} and eROSITA \citep{Merloni2012} surveys are indicated by green and black lines, respectively. The gray dotted line indicates the Milky Way at Galactic latitude $b = 0^\circ$.}
\label{Fig:observed_sky}
\end{figure*}

The first-generation blind \hi surveys, including the \hi Parkes All-Sky Survey (HIPASS; \citealt{Barnes2001}) and the Arecibo Legacy Fast ALFA Survey (ALFALFA; \citealt{Haynes2018}), provided foundational catalogs of \hi-selected galaxies in the local Universe. HIPASS covered $\sim$30\,000\,deg$^{2}$ but was limited to $z < 0.042$ and only detected about $5\,000$ \hi targets with modest sensitivity ($\sigma_{\rm rms} \sim 13$\,mJy\,beam$^{-1}$; \citealt{Koribalski2004,Meyer2004,Wong2006}), while ALFALFA significantly improved sensitivity ($\sigma_{\rm rms} \sim 2.4$\,mJy\,beam$^{-1}$) and angular resolution ($\sim$3$\dotmin$5), cataloging $\sim$31\,500 sources out to $z \sim 0.06$ over $\sim$7\,000\,deg$^{2}$ \citep{Haynes2018}. Despite their success, these surveys were limited by shallow depth, incomplete sky coverage, and insufficient sensitivity to low-\hi-mass systems ($M_{\rm \hj} \lesssim 10^{8}\,M_{\odot}$), leaving the intermediate-redshift regime ($0.1 \lesssim z \lesssim 0.4$) largely unexplored by wide-area \hi observations.

\begin{figure}[H]
\centering
\includegraphics[width=0.48\textwidth, angle=0]{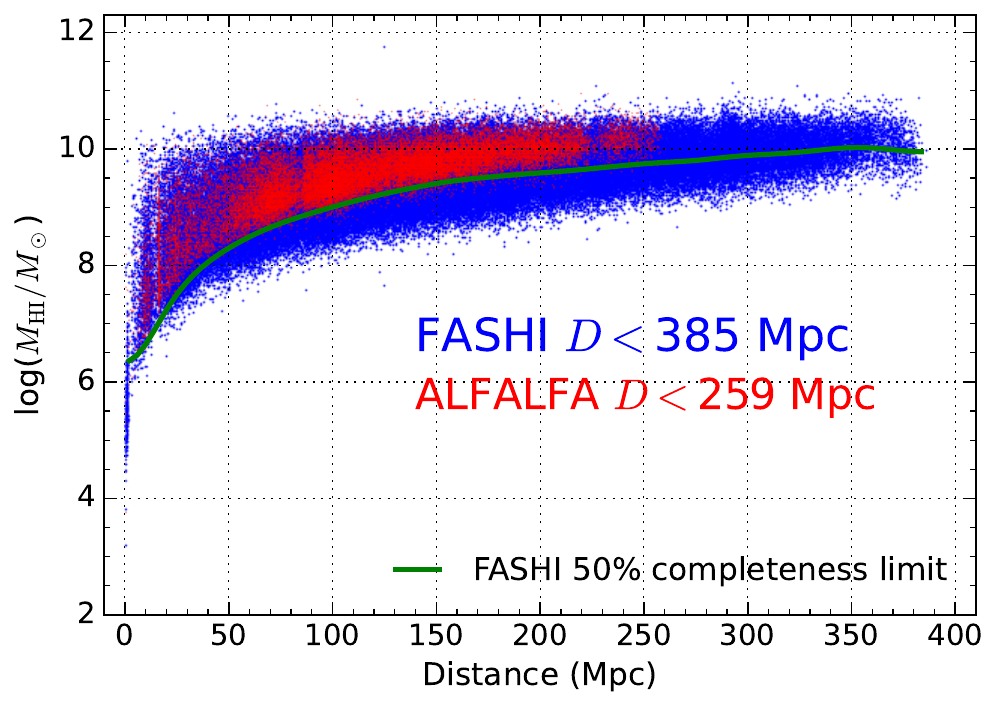}
\caption{Mass-distance diagram for the FASHI \hi\ sources (blue points, see Table \ref{tab:exgalcat}), compared with the ALFALFA $\alpha$100 sample (red points, \citealp{Haynes2018}). The green line traces the FASHI 50\% completeness limit, measured as the average $S_{\hj,50\%}$ at each distance bin (See text for details).}
\label{Fig:mass_distance}
\end{figure}

The Five-hundred-meter Aperture Spherical radio Telescope (FAST), the world's most sensitive single-dish radio telescope, addresses this need through the FAST All-Sky \hi survey (FASHI). FASHI has covered $\sim$19\,500\,deg$^{2}$ ($-14^{\circ} < {\rm DEC} < +66^{\circ}$) over the 1.0--1.5\,GHz band with a median sensitivity of $\sim$0.5\,mJy\,beam$^{-1}$ at 6.4\,km\,s$^{-1}$ spectral resolution, an order of magnitude deeper than ALFALFA over a substantially larger area. With its unprecedented combination of sensitivity and survey volume, FASHI enables robust constraints on the low-mass slope (see Figure\,\ref{Fig:mass_distance}) of the \hi mass function (HIMF), precise measurements of the cosmic \hi density, $\Omega_{\rm HI}$ in complementary stacking studies, and mapping of large-scale \hi structure over a cosmologically representative volume \citep{Giovanelli2015}. The survey is also sensitive to faint and rare populations, including low-surface-brightness dwarf galaxies \citep{Janowiecki2015}, nearly dark galaxies \citep{Giovanelli2013,Xu2023dark,Monaci2026}, diffuse intergalactic gas \citep{Zhu2021, Liu2023}, and Galactic high-velocity clouds \citep{Adams2013,Liu2025hvc}.

Beyond its stand-alone scientific value, FASHI provides a unique foundation for multi-wavelength synergy. By combining the deep FASHI 21\,cm observations \citep{Zhang2024} with spectroscopic redshifts from the Dark Energy Spectroscopic Instrument (DESI), which covers $\sim$17\,000\,deg$^{2}$ \citep{desi2025}, a joint optical--radio sample spanning $\sim$12\,000\,deg$^{2}$ over $0 < z < 0.41$ can be constructed \citep{ZhangNA2026}. The overlap between FASHI and the eROSITA X‑ray survey \citep{Merloni2012} similarly exceeds $\sim$12\,000\,deg$^{2}$ (see Figure\,\ref{Fig:observed_sky}), enabling future studies of the baryon cycle by combining cold \hi gas with hot X‑ray emitting circumgalactic and intracluster gas. As the most powerful wide-area \hi survey in the pre-SKA era \citep{Blyth2015}, FASHI serves as a critical foundation for future joint analyses with interferometric facilities, enabling detailed studies of \hi morphology, kinematics, and the baryon cycle in the nearby Universe \citep{Koribalski2020}.

While FASHI excels in wide-area statistical studies, recent interferometric \hi surveys have pushed to higher redshifts and finer angular resolution. Programs using the Westerbork Synthesis Radio Telescope with Apertif \citep{Adams2019}, the Australian Square Kilometre Array Pathfinder (ASKAP) for the Widefield ASKAP L-band Legacy All-sky Blind survey (WALLABY; \citealt{Koribalski2020}), the Karl G. Jansky Very Large Array (VLA) for the COSMOS \hi Large Extragalactic Survey (CHILES; \citealt{Fernandez2016,Luber2025}), the MeerKAT radio telescope for the \hi component of the MeerKAT International Gigahertz Tiered Extragalactic Exploration (MIGHTEE) survey \citep{Maddox2021,Jarvis2025,Heywood2024}, and the Looking At the Distant Universe with the MeerKAT Array (LADUMA) project \citep{Kazemi-Moridani2025} have reached redshifts of $z \gtrsim 0.3$. However, these surveys typically cover areas $\lesssim 100\,\mathrm{deg}^2$, making them susceptible to cosmic variance, particularly at low redshifts. At $z\gtrsim0.3$, the same surveys probe larger comoving volumes, partially mitigating this concern. This trade-off between depth and sky coverage highlights the complementary nature of single-dish and interferometric approaches.

Single-dish and interferometric \hi surveys offer distinct yet complementary strengths. The unparalleled sensitivity and survey speed of FAST enable FASHI to construct the largest statistical sample of \hi-selected galaxies in the local Universe, providing robust measurements of the \hi mass function and cosmic \hi density. However, the large beam of FAST ($\sim$2$\dotmin$9) limits its ability to resolve internal gas kinematics or to unambiguously associate \hi emission with individual optical counterparts in crowded regions. Interferometric surveys such as MIGHTEE, CHILES, and LADUMA complement these single-dish efforts by delivering high spatial resolution ($\sim$arcseconds), enabling detailed studies of \hi morphology, kinematics, and the connection to multi-wavelength data without source confusion. Conversely, interferometers are less sensitive to extended, diffuse \hi emission and typically cover smaller areas. The synergy between these approaches is particularly powerful: FASHI identifies the statistical sample and detects diffuse gas over wide areas, while interferometric follow-up resolves the detailed properties of individual systems. Notably, the COSMOS field observed by MIGHTEE-HI \citep{Heywood2024} overlaps with FAST observations \citep{Pan2024}, providing a direct opportunity for such synergistic studies.

In this paper, we present the second data release (DR2) of the FASHI project, covering sources with redshifts $z < 0.09$ in the local Universe. Section~\ref{sec:survey} describes the survey design and observing strategy. Section~\ref{sec:obs_setup} details the observational setup. Section~\ref{sec:data_reduc} outlines the data reduction procedures, including baseline subtraction and source extraction. The released FASHI catalog of extragalactic \hi sources and the source characterization are presented in Section~\ref{sec:catalog}, along with comparisons to ALFALFA, DESI, and SDSS sources. Section~\ref{sec:discuss} assesses the catalog’s reliability and completeness, presents the \hi mass function (HIMF), and discusses caveats for its use. A summary is provided in Section~\ref{sec:summ}. Throughout this work, we assume a flat $\Lambda$CDM cosmology with $H_{0} = 70\,\kms {\rm Mpc}^{-1}$, $\Omega_{\mathrm{M}} = 0.3$, and $\Omega_{\Lambda} = 0.7$.

\section{The FASHI Survey Strategy}
\label{sec:survey}

FAST is currently the most sensitive single-dish radio telescope in the world. Located in Guizhou, China (25\degree 39\arcmin 11\arcsec\,N, 106\degree 51\arcmin 24\arcsec\,E), FAST is capable of observing sources within a maximum zenith angle of $40^{\circ}$ \citep{Nan2011, Jiang2019, Jiang2020}. FASHI survey is designed to map the entire sky accessible to FAST, covering declinations from $-14^{\circ}$ to $+66^{\circ}$ in the frequency range 1.0--1.5\,GHz, with high sensitivity and adequate spatial and spectral resolution \citep{Zhang2024}.

Between August 2020 and July 2025, FASHI observed approximately $19\,500$\,deg$^{2}$, corresponding to $47.3\%$ of the full sky ($41\,252.96$\,deg$^{2}$). Within this area, the survey has identified about $156\,000$ extragalactic \hi sources at redshifts $z < 0.09$, yielding an average number density of $\sim$8.0 sources per square degree. The key technical parameters of the FASHI survey are summarized in Table~\ref{tab:obs}.

The FASHI survey adopts a flexible drift-scan observing strategy, optimized for FAST’s schedule-filler observing mode. Observations are carried out at fixed declinations, enabling an efficient blind search for \hi emission while maximizing the utilization of available telescope time. This strategy naturally leads to non-uniform sky coverage, with some regions observed multiple times and thus reaching greater sensitivity.

\section{Observational setup}
\label{sec:obs_setup}

FASHI utilizes FAST's 19-beam receiver array in drift-scan mode for efficient wide-area coverage. The spectral backend records dual linear polarizations (XX/YY) across 1.0--1.5\,GHz, with 64k channels yielding a native resolution of 7.63\,kHz ($\sim$1.6$\rm\,km\,s^{-1}$ at 1.4\,GHz). Data are recorded at 1\,second integrations. The telescope's 500\,m aperture (effective $\sim$$300$\,m) provides a half-power beamwidth of $2\dotmin9$ at 1.4\,GHz and a pointing accuracy better than $15''$. Intensity calibration employs periodic 1-second noise-diode injections ($\sim$11\,K) every 32 or 64 seconds, with a beam-dependent gain of 13--17\,K\,Jy$^{-1}$ at 1400\,MHz \citep{Nan2011,Jiang2019,Jiang2020,Han2021}. The systematic uncertainty of the FAST flux calibration is better than 5\% \citep{Jing2024}. Key observational parameters are summarized in Table\,\ref{tab:obs}.

\begin{table}[H]
\caption{\textbf{FASHI DR2 technical details.}}
\label{tab:obs}
\vskip 5pt
\centering
\footnotesize{
\begin{tabular}{l|l}
\hline
\hline
Parameters                      & Values \\
\hline
RA range                        & $\rm 0^h\leq RA \leq24^h$  \\
DEC range                       & $\rm -14\degree \lesssim DEC \lesssim +66\degree$  \\
Receiver                        & 19-beam array \\
Polarizations per beam          & 2 linear (XX, YY) \\
Beam size (FWHM)                & 2$\dotmin$9 at 1420\,MHz \\
Gain                            & $13\sim17$\,K\,Jy$^{-1}$ \\
$T_{\rm{sys}}$                  & $16\sim19$\,K \\
Total frequency coverage        & $1050\sim1450$\,MHz \\
Released frequency range        & $1305.7\sim1422.8$\,MHz \\
c$z_{\odot}$ range              & $-510.0\sim26320.6\,\kms$\ \\
Redshift $z$ range              & $<0.0878$ \\
Spectral channels               & 65536 (before smoothing) \\
Channel spacing                 & 7.629\,kHz or $1.6\,\kms$\ at 1420\,MHz \\
Spectral resolution             & $6.4\,\kms$\ (after smoothing) \\
Median rms $\sigma_{\rm rms}$            & $\sim$1.04\,mJy at $6.4\,\kms$\ resolution \\
Median sensitivity $f_\sigma$          & $\sim$0.57\,$\mjyb$ at $6.4\,\kms$\ resolution \\
Coverage area                   & 19\,500~deg$^{2}$ \\
Source number                   & 156\,411 \\
\hline
\hline
\end{tabular}}
\begin{flushleft}
\textbf{Notes.} The definitions of the spectral rms $\sigma_{\mathrm{rms}}$ and the per-source detection sensitivity $f_\sigma$ are provided in Column\,9 of Section\,\ref{sec:catalog_col}. \\
\end{flushleft}
\end{table}

The primary observing mode for FASHI is the drift-scan \texttt{DecDriftWithAngle} mode utilizing FAST's 19-beam receiver array \citep{Li2018,Jiang2019,Jiang2020}. This mode minimizes system complexity and self-generated RFI while maintaining stable pointing and full system gain. During observations, the azimuth arm is fixed on the meridian at predefined J2000 declinations, with successive scans spaced by $21\dotmin65$. The feed array is rotated by $23\dotdeg4$ to achieve super-Nyquist sampling along the drift tracks, ensuring full spatial sampling within the beam.

To supplement coverage in regions missed by the primary drift-scan strategy, FASHI also employs the \texttt{MultiBeamOTF} (on-the-fly) scanning mode \citep{Jiang2020}. This mode is used during schedule-filler time to target specific declination strips that were not uniformly sampled in drift scans, thereby increasing overall sky coverage. The instrumental setup in this mode is identical to that used in \texttt{DecDriftWithAngle} observations, except for the controlled scanning in declination.

\section{Data reduction}\label{sec:data_reduc}

Detailed data reduction procedures are described in the FASHI first data release paper \citep{Zhang2024}. Here we provide only a brief summary.

\subsection{Spectral data reduction with \texttt{HiFAST}}

The FASHI data were processed using the dedicated FAST pipeline \texttt{HiFAST} \citep{Jing2024}. This pipeline integrates modules for antenna-temperature correction, baseline fitting, radio frequency interference (RFI) mitigation \citep{Zhang2022}, standing-wave removal \citep{Xu2025}, image gridding, flux calibration \citep{Liu2024}, and data cube generation. A key step is the baseline correction: each original spectrum was processed using the asymmetrically reweighted penalized least squares algorithm \citep[\texttt{arPLS};][]{Baek2015}, where strong emission or absorption lines exceeding $3\sigma$ were masked prior to the final baseline fit. For the present release, all spectra were smoothed to a final velocity resolution of $6.4\,\kms$\,per\,channel. The resulting data cubes were gridded at a pixel scale of $1'$, satisfying the Nyquist sampling criterion. Flux calibration was performed using the zenith-angle-dependent gain curves established for the 19-beam receiver \citep{Jiang2019}, which are applicable for observations with zenith angles $\mathrm{ZA} \lesssim 40\degree$.

\subsection{Source extraction with \texttt{SoFiA}}
\label{sec:sofia}
Source extraction for FASHI employed a two-step approach combining automated detection with visual validation. Initial candidates were identified using version 2 of the \hi Source Finding Application \citep[\texttt{SoFiA};][]{Serra2015,Westmeier2021,Westmeier2022}, which performs multi-scale smoothing across spatial and spectral dimensions, estimates local noise in each smoothed iteration, and selects pixels exceeding a user-defined signal-to-noise threshold. For FASHI, a detection threshold of $4.5\sigma$ (where $\sigma$ is the local noise within the source bounding box) was adopted to generate the preliminary source list. This threshold was chosen empirically: a lower threshold ($4\sigma$) produced an excessive number of false positives, significantly increasing the visual inspection workload, while a higher threshold ($5\sigma$) rejected a noticeable fraction of real sources that were later confirmed by optical spectroscopy. The $4.5\sigma$ threshold was found to provide the optimal trade-off between completeness and reliability. All candidates were then inspected interactively using moment maps and integrated profiles, supplemented by available optical counterparts, to reject false detections and refine source boundaries, ensuring a reliable final catalog.

\subsection{Optically guided source extraction}
\label{sec:opt_guide}

While \texttt{SoFiA} provides robust automated detection, it occasionally misses faint \hi sources. Given the substantial sky overlap of more than 12\,000 deg$^2$ between FASHI and the DESI DR1 and SDSS spectroscopic surveys \citep{desi2025,Abazajian2009}, we leverage these optical data to complement our source extraction and improve sample completeness.

We adopt the following procedure for optically guided extraction. First, we construct a candidate list from the DESI and SDSS catalogs using tight matching tolerances of $\delta_{\mathrm{RA}} \leq 1'$, $\delta_{\mathrm{DEC}} \leq 1'$, and $\delta_{\mathrm{velocity}} \leq 30\,\kms$. For each optical counterpart, we extract the corresponding \hi spectrum and require an emission signal exceeding $3\sigma$ for detection. We then merge the \hi sources identified by this method with those found by \texttt{SoFiA}. Duplicate sources are removed using a slightly larger tolerance of $\delta_{\mathrm{RA}} \leq 2'$, $\delta_{\mathrm{DEC}} \leq 2'$, and $\delta_{\mathrm{velocity}} \leq 50\,\kms$. This hybrid approach significantly increases the detection completeness, particularly for low-SNR \hi sources with secure optical redshifts. A similar methodology has been employed by \citet{Jarvis2025} to search for high-redshift \hi galaxies in the MIGHTEE survey, where DESI spectroscopic redshifts were used to guide the extraction of 21\,cm emission.

\section{The FASHI catalog and Analysis}
\label{sec:catalog}

\subsection{Extragalactic \hi catalog}
\label{sec:catalog_col}

\begin{table*}
\caption{\textbf{FASHI DR2 Extragalactic \hi Source Catalog}}
\label{tab:exgalcat}
\vskip 5pt
\centering \small  
\setlength{\tabcolsep}{1.2mm}{
\begin{tabular}{cccccccccccccccc}
\hline \hline
[1]  &  [2]  & [3]   & [4]  & [5] & [6]  & [7]  & [8] & [9] & [10] & [11] & [12] & [13] & [14] & [15]  \\ 
FASHI ID &  RA & DEC &   c$z_{\odot}$ &  ell$_{\rm maj,\,min,\,pa}$ & W$_{50}$ & W$_{20}$ & $F_{\rm peak} $& $\sigma_{\rm rms}$  & $S_\hj$   & SNR & $D$ & log$M$ & $C$ & $V_{\rm max}$ \\
 & deg & deg   & $\kms$    & $~~',~~',~~^\circ$   & $\kms$     & $\kms$     & mJy & mJy& mJy$\cdot\kms$ &  & Mpc & $M_{\odot}$  &  & Mpc$^3$   \\
\hline
20260000001&0.003&5.443&12047.9&3.5,\,3.5,\,0&231.3&426.9&7.3&1.4&1896.9&25.8&164.9&10.1&0.901&$2.27\times10^{7}$\\ 
20260000002&0.004&24.909&11190.3&5.0,\,5.0,\,0&295.7&338.3&10.7&1.7&2428.8&22.8&149.5&10.1&0.915&$1.93\times10^{7}$\\ 
20260000003&0.004&32.702&10592.8&2.9,\,2.9,\,0&214.9&233.1&3.7&1.1&453.4&8.2&142.4&9.3&0.139&$2.14\times10^{6}$\\ 
20260000004&0.005&15.892&5995.5&2.7,\,2.7,\,0&173.6&223.8&4.0&1.3&645.2&10.5&80.4&9.0&0.274&$6.05\times10^{5}$\\ 
20260000006&0.006&47.273&5002.1&5.2,\,5.2,\,0&427.6&446.8&26.7&2.9&6607.2&30.3&63.8&9.8&0.916&$1.80\times10^{6}$\\ 
20260000008&0.010&23.085&4451.4&4.8,\,4.8,\,0&164.7&180.4&14.7&2.1&1920.2&20.4&54.9&9.1&0.965&$1.28\times10^{6}$\\ 
20260000009&0.010&34.524&12699.0&2.8,\,2.8,\,0&238.3&278.7&3.8&1.7&901.0&9.8&173.9&9.8&0.208&$4.70\times10^{6}$\\ 
20260000011&0.012&35.805&9533.2&4.2,\,4.2,\,0&167.0&200.7&4.3&1.0&536.5&11.2&127.7&9.3&0.726&$5.99\times10^{6}$\\ 
20260000012&0.014&26.016&10419.4&5.0,\,5.0,\,0&286.8&317.9&14.5&2.7&2572.2&15.9&137.0&10.0&0.796&$9.33\times10^{6}$\\ 
20260000013&0.022&36.757&4870.7&2.9,\,2.9,\,0&72.3&164.7&3.1&0.6&278.5&15.6&57.7&8.3&0.836&$7.52\times10^{5}$\\ 
... ... \\
\hline
\end{tabular}}
\begin{flushleft}
\textbf{Notes.} The complete catalog of 156\,411 sources is provided in the online supplementary material and at \url{https://fast.bao.ac.cn/cms/article/271/} and \url{https://zcp521.github.io/fashi}, with additional parameter columns available for download.
\\
\end{flushleft}
\end{table*}

Table\,\ref{tab:exgalcat} summarizes the key parameters of the FASHI DR2 extragalactic \hi source catalog. The complete catalog, comprising 156\,411 sources with derived parameters, is available in the online supplementary material and at \url{https://fast.bao.ac.cn/cms/article/271/} and \url{https://zcp521.github.io/fashi}. The full version provides high-precision information including: J2000 centroid coordinates in \texttt{Jhhmmss.ss$\pm$ddmmss.s} format, spectral line central frequency, radio-defined velocity, integrated velocity range, detection sensitivity, sample completeness, maximum comoving volume, associated uncertainties and so on. Main parameters are described below, and parameters exclusive to the online table are also noted accordingly.

\begin{itemize} 

\item Column\,1: FASHI source ID. A unique sequential identifier assigned to each extragalactic \hi source in the catalog.

\item Columns\,2--3: Right ascension and declination (J2000). Celestial coordinates (RA, DEC in degrees) of the \hi brightness centroid for each source. The corresponding sexagesimal format \texttt{Jhhmmss.ss±ddmmss.s} is provided in the table. The positional uncertainty listed online is approximated as the beam size ($\sim$2$\dotmin9$) divided by the source’s \hi signal‑to‑noise ratio \citep{Koribalski2004}.

\item Column\,4: Heliocentric velocity. Optical-definition heliocentric velocity $cz_{\odot}$ in $\kms$, converted from the radio convention ($\delta\nu/\nu$). Its uncertainty is $\sigma(cz_{\odot}) = 3\sqrt{P\,\delta v}/{\rm SNR}$, where $\delta v=6.4$\,$\kms$\ is the spectral resolution and $P = (W_{20}-W_{50})/2$ measures the profile edge steepness \citep{Fouque1990,Koribalski2004}. The corresponding central frequency, radio velocity, redshift $z_{\odot}$ and their uncertainties are listed in the online table. The $cz_{\odot}$ distribution of all sources is shown in Figure\,\ref{Fig:hist_fast}.

\item Column\,5: Ellipse parameters $ell_{\rm maj,\,min,\,pa}$, defining the adaptive aperture for flux integration \citep[see details in][]{Zhang2024}. The major and minor axes are in arcmin, and the position angle in degrees. This elliptical extraction area is optimized to recover the \hi flux and does not trace the galaxy's intrinsic morphology.

\item Column\,6: Velocity width $W_{50}$ (in $\kms$), measured at 50\% of each peak flux using busy-function fitting. The uncertainty is $\sigma(W_{50}) = 2\sigma({\rm c}z_{\odot})$. The distribution of $W_{50}$ values is presented in Figure\,\ref{Fig:hist_fast}.

\item Column\,7: Velocity width $W_{20}$ (in $\kms$), measured at 20\% of each peak flux using busy-function fitting. The uncertainty is $\sigma(W_{20}) = 3\sigma({\rm c}z_{\odot})$. The $W_{20}$ distribution appears in Figure\,\ref{Fig:hist_fast}.

\item Column\,8: Peak \hi line flux $F_{\rm peak}$ (in mJy), measured at the maximum of the integrated spectrum via busy‑function fitting. Its uncertainty is derived from the rms noise $\sigma_{\rm{rms}}$ given in Column\,9.

\item Column\,9: Integrated spectral noise $\sigma_{\mathrm{rms}}$ (mJy), measured from signal- and RFI-free channels at $6.4\,\kms$ resolution. Its distribution is shown in Figure\,\ref{Fig:hist_fast}. The per-source detection sensitivity $f_{\sigma}$ ($\mjyb$) is derived from $\sigma_{\mathrm{rms}}$ as
\begin{equation}
f_{\sigma} = \frac{2\sigma_{\mathrm{rms}}\Omega_{\mathrm{beam}}}{\theta_{\mathrm{beam}}\sqrt{ell_{\mathrm{maj}}ell_{\mathrm{min}}}} = \frac{\pi\sigma_{\mathrm{rms}}\theta_{\mathrm{beam}}}{2\ln(2)\sqrt{ell_{\mathrm{maj}}ell_{\mathrm{min}}}},
\end{equation}
where $\Omega_{\mathrm{beam}}$ is the beam solid angle, $\theta_{\mathrm{beam}}$ the beam size, and $ell_{\mathrm{maj}}$, $ell_{\mathrm{min}}$ the source aperture dimensions (Column\,5). The median $f_{\sigma}$ is $\sim0.57\,\mjyb$, with the full value listed in the online table. The spatial variation of sensitivity, shown in Figure\,\ref{Fig:sensitivity}, reflects the non-uniform coverage of the schedule-filler observing mode.

\begin{figure}[H]
\centering
\includegraphics[width=0.48\textwidth, angle=0]{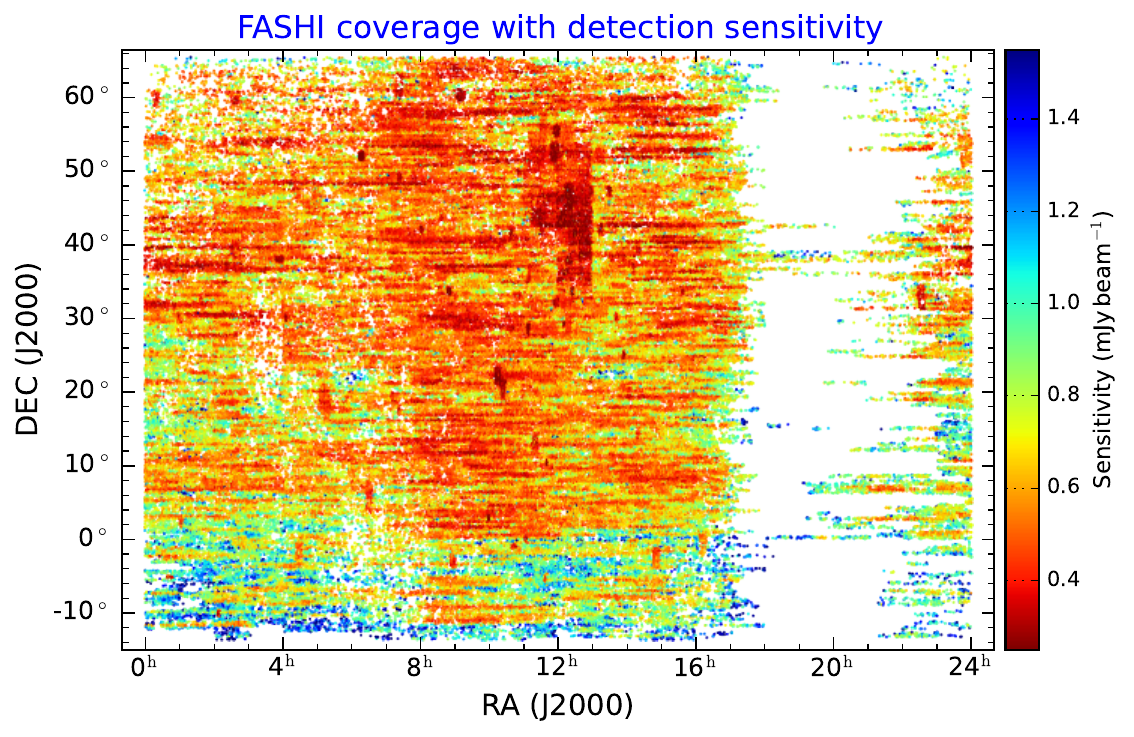}
\caption{Distribution of the detection sensitivity, $f_\sigma$, calculated as the rms noise per beam, in the FASHI survey coverage. The median $f_\sigma$ is $\sim$0.57\,$\mjyb$. The effect of non-uniform $f_\sigma$ is taken into account by the sensitivity-dependent completeness.}
\label{Fig:sensitivity}
\end{figure}

\begin{figure*}
\centering
\includegraphics[width=0.49\textwidth, angle=0]{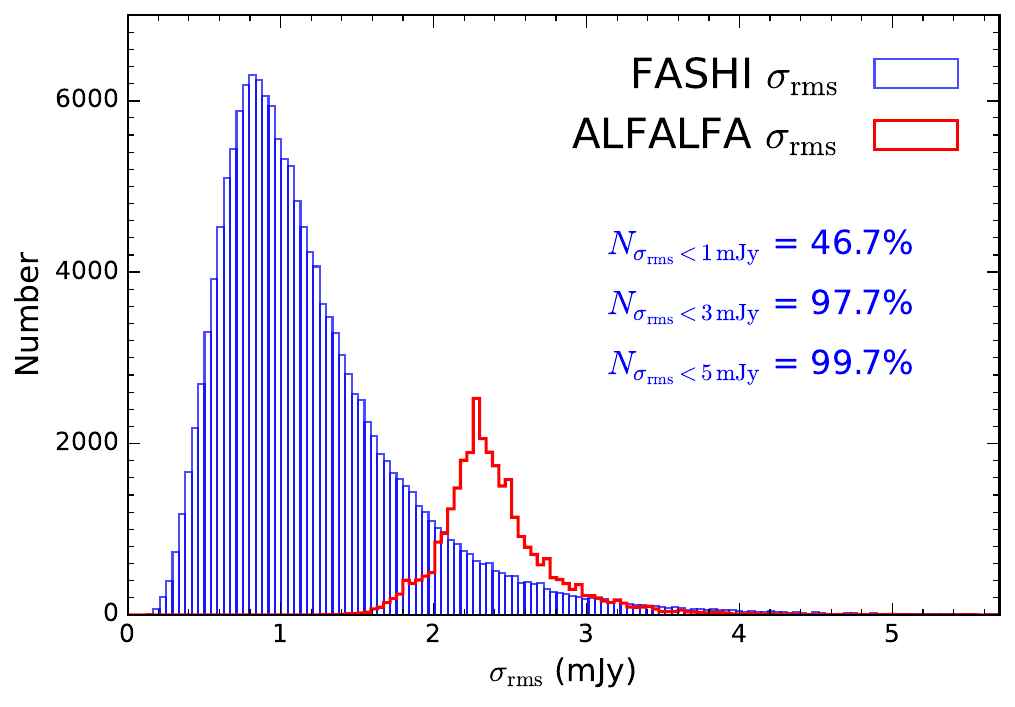}
\includegraphics[width=0.49\textwidth, angle=0]{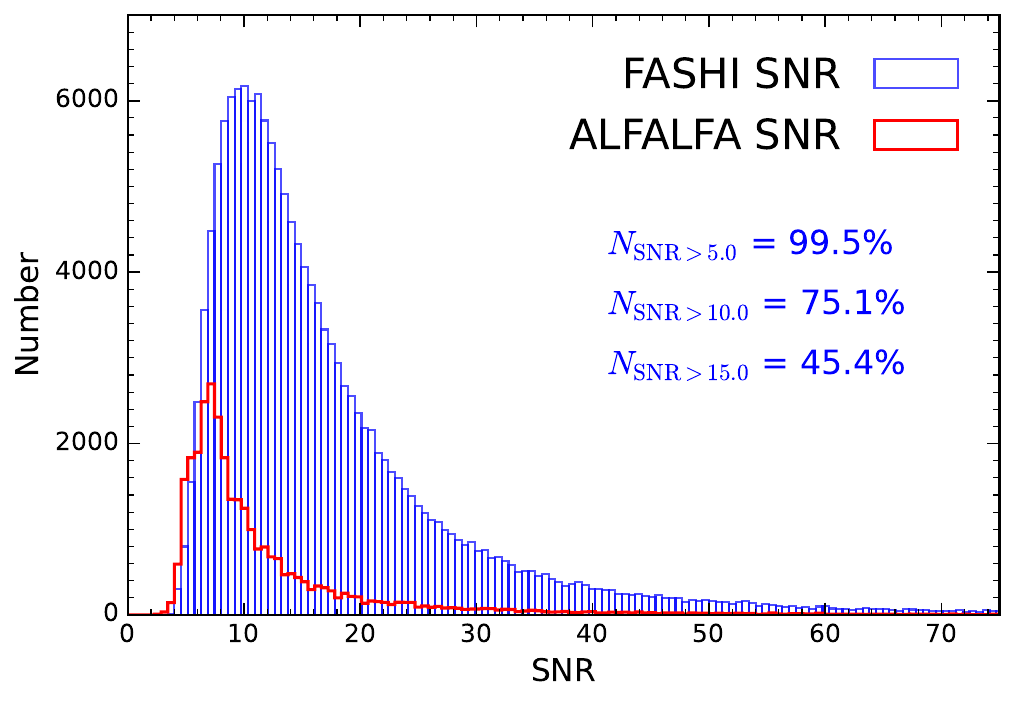}
\includegraphics[width=0.49\textwidth, angle=0]{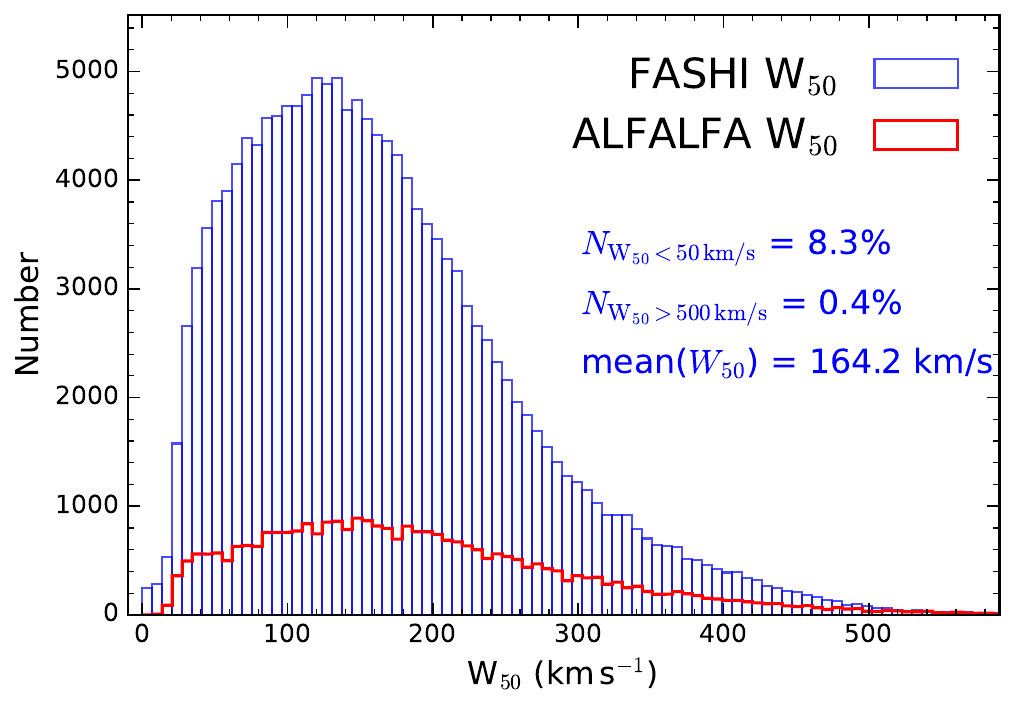}
\includegraphics[width=0.49\textwidth, angle=0]{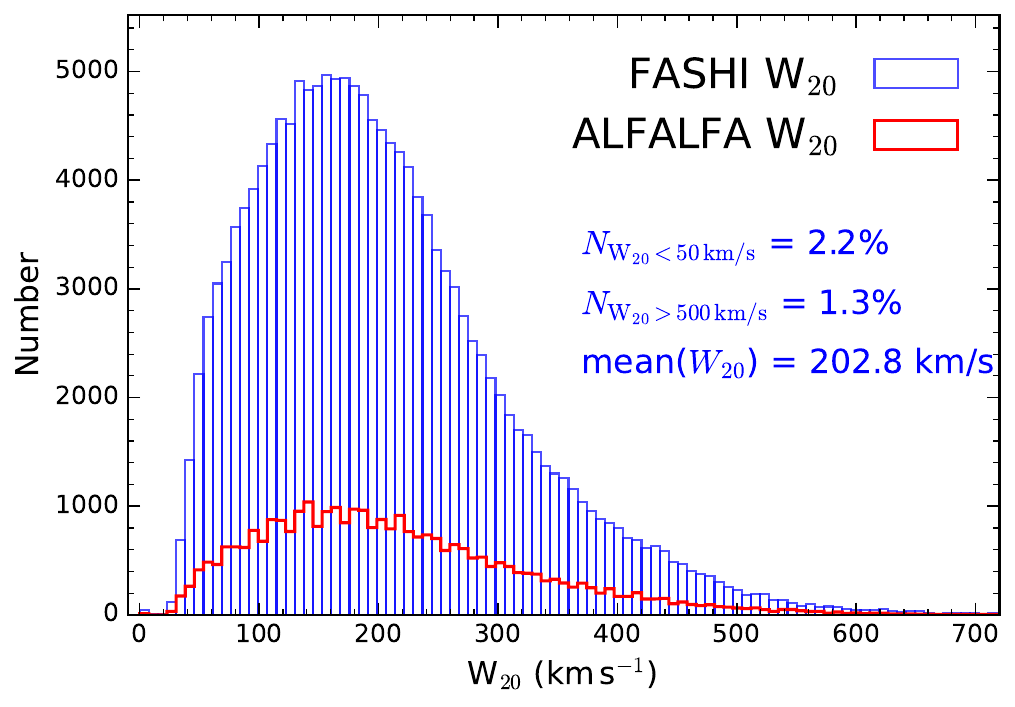}
\includegraphics[width=0.49\textwidth, angle=0]{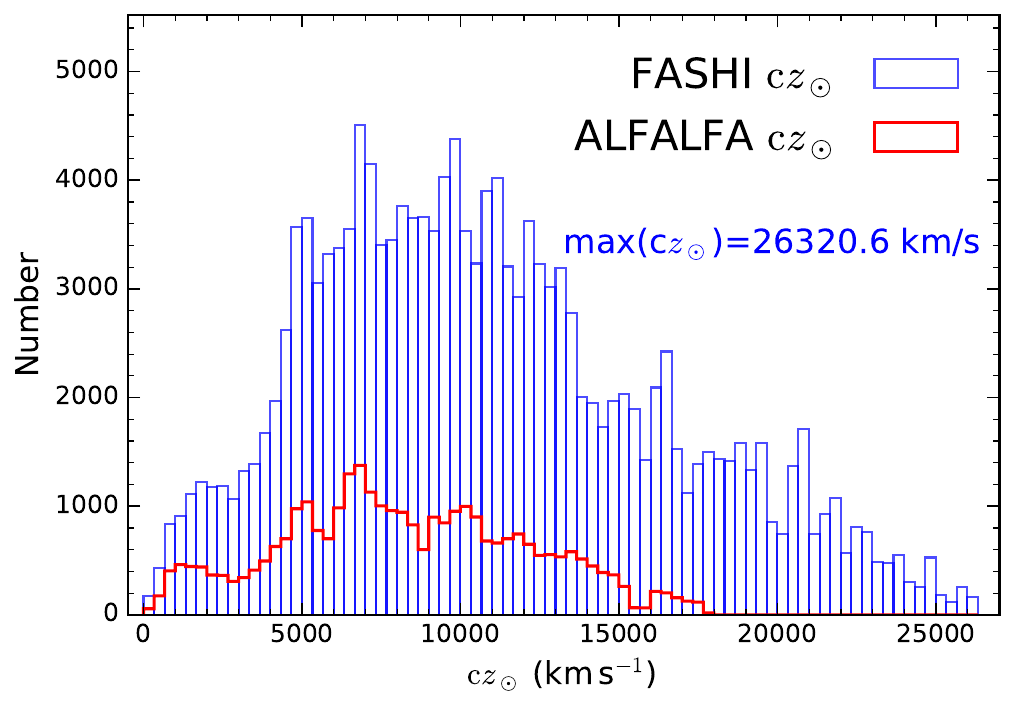}
\includegraphics[width=0.49\textwidth, angle=0]{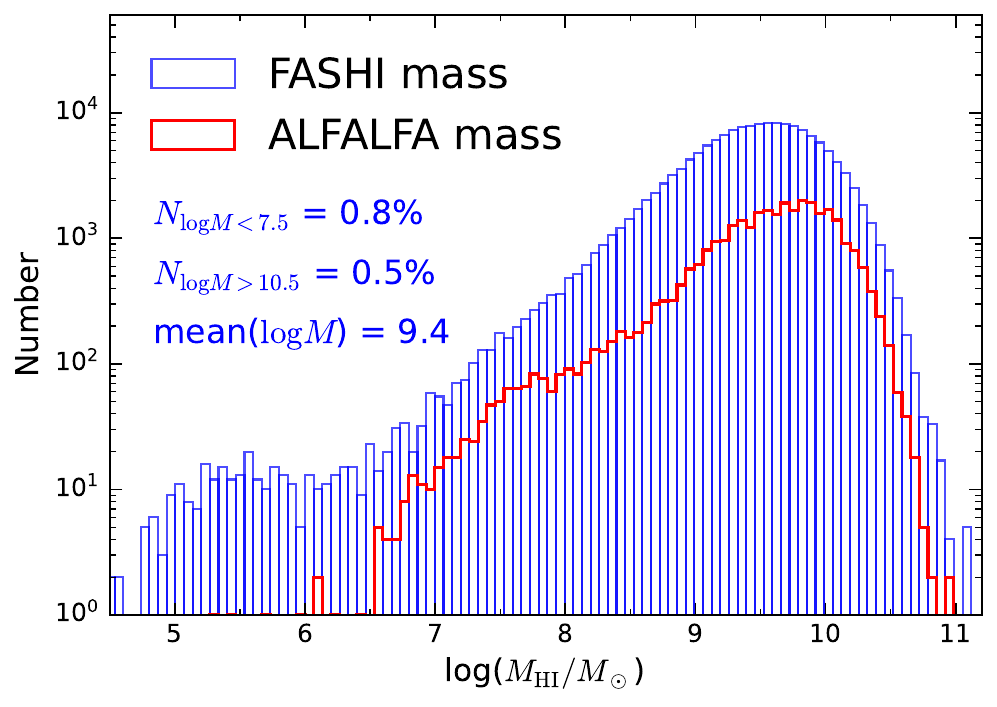}
\caption{Distributions of measured properties including spectral rms ($\sigma_{\rm rms}$), SNR, line widths ($W_{50}$, $W_{20}$), heliocentric velocity ($cz_{\odot}$), and \hi mass, presented for the FASHI and ALFALFA catalogs. Statistical parameters for FASHI are shown in each panel.}
\label{Fig:hist_fast}
\end{figure*}

\item Column\,10: Integrated \HI flux density $S_{\rm bf}$ (in $\mjybkms$) from busy‑function fitting. The signal‑to‑noise ratio (Column\,11) and \HI mass (Column\,13) are derived from $S_{\rm bf}$ here. Its uncertainty is $\sigma(S_{\rm bf}) = \sqrt{N}\,\delta v\,\sigma_{\rm rms}$, where $N$ is the number of channels containing the \HI line and $\delta v=6.4\,\kms$. Integrated \hi flux density $S_{\rm sum}$ from direct summation of signal channels within the velocity range are provided in the online table. It shares the same uncertainty level as $S_\hj$. For a fraction of sources, $S_\hj$ exceeds $S_{\rm sum}$; this systematic difference likely arises from the combined influence of signal‑to‑noise, baseline quality, integration boundaries, and spectral‑profile shape.

\item Column\,11: Signal-to-noise ratio (SNR). The SNR is defined as
   \begin{equation}
	{\rm SNR} = \left( \frac{S_\hj}{W_{50}} \right) \frac{w_{\rm smo}^{1/2}}{\sigma_{\rm rms}},
	\label{eq:eqsn}
	\end{equation}
where $S_\hj$ is in $\mjybkms$, $w_{\rm smo}=W_{50}/6.4$ is the smoothing width in channels of $6.4\,\kms$, and $\sigma_{\rm rms}$ is the rms noise (mJy) per channel. This definition yields the peak signal-to-noise ratio of the H\,{\sc i} line, rather than the integrated SNR. We adopt this definition for consistency with previous surveys (e.g., \citealt{Haynes2018}). The SNR distribution is shown in Figure\,\ref{Fig:hist_fast}.

\item Column\,12: Distance $D$ (Mpc). Distances are primarily derived from the Cosmicflows-4 (CF4) online distance calculator \citep{Kourkchi2020,Valade2024}, which converts galaxy redshifts into expectation distances using a model of the local velocity field. This flow model accounts for large-scale peculiar motions. However, for sources in cluster environments, it does not remove the effect of large internal velocity dispersions, and the resulting distances remain subject to substantial uncertainties \citep{Haubner2025}. For sources not covered by CF4, we adopt individual distances from \citet{Karachentsev2013} catalog of 869 nearby galaxies within $11$\,Mpc. The remaining sources without a reliable distance estimate are assigned a placeholder value of $1.0$\,Mpc with a conservative $50\%$ uncertainty. The distance uncertainty is assigned based on the source environment and redshift \citep{Stiskalek2026}. For field galaxies at $z\gtrsim0.02$ ($D\gtrsim86$\,Mpc), the uncertainty is $5\%$, dominated by the Hubble constant uncertainty. For field galaxies at $0.01<z<0.02$ ($11\lesssim D\lesssim86$\,Mpc), we adopt a $7.5\%$ uncertainty to account for residual peculiar velocities. For nearby galaxies with $D\lesssim11$\,Mpc, we adopt a $25\%$ uncertainty. For galaxies in the Virgo cluster, we assign a $39\%$ uncertainty, following the Virgo Zone of Influence scheme of \citet{Haubner2025} for the CF4 flow model. For the Coma cluster, we assign an $11\%$ uncertainty, following the general error scheme of \citet{Haubner2025} at a distance of approximately $100$\,Mpc. The distance distribution is shown in Figure\,\ref{Fig:mass_distance}.

\item Column\,13: Logarithmic \hi mass $\log (M_\hj/M_\odot$), computed as
   \begin{equation}
	\frac{M_\hj}{M_\odot} = \frac{2.356\times 10^{5}}{1+z_{\odot}}\,\left(\frac{D}{\mathrm{Mpc}}\right)^{2}\frac{S_\hj}{\rm Jy\,km\,s^{-1}},
	\label{eq:himass}
	\end{equation}
where $D$ is the distance (Column\,12), $S_\hj$ the integrated flux (Column\,10), and $1+z_{\odot}$ the cosmological correction factor \citep{Haynes2018}. The uncertainty is propagated from the distance and flux errors. The mass histogram appears in Figure\,\ref{Fig:hist_fast}.

\item Column\,14: Sample completeness $C$, which quantifies the fraction of galaxies detected by FASHI at a given integrated flux density (see Eq.\,(\ref{eq:comp})). This correction factor must be applied to account for missing objects when measuring the total number density of \hi-selected galaxies.

\item Column\,15: The maximum comoving volume $V_{\mathrm{max}}$ within which a galaxy could still be detected, given its \hi mass and the survey's flux limit. The calculation of the comoving volume is measured with the survey area of $19482\deg^2$ and the maximum detectable distance.
\end{itemize}

\subsection{Comparison with ALFALFA sources}
\label{sec:match_alfa}

Figure\,\ref{Fig:observed_sky} compares the sky distributions of \hi sources from FASHI and ALFALFA. FASHI has surveyed a significantly larger area than ALFALFA, detecting $156\,411$ sources over $19\,482\deg^2$ ($\sim$$8.0\,\deg^{-2}$), compared to ALFALFA’s $31\,502$ sources over $6\,518\,\deg^2$ ($\sim$$4.8\,\deg^{-2}$) based on our calculations. As shown in the \hi mass–distance diagrams (Figure\,\ref{Fig:mass_distance}), FASHI reaches both more distant and lower‑mass systems than ALFALFA (see also Figures~\ref{Fig:hist_fast} and \ref{Fig:polar}), demonstrating the superior sensitivity of FAST. The broad sky and frequency coverage of FASHI further enable multi‑wavelength studies of galaxies by combining with DESI and eROSITA data, providing powerful constraints on galaxy formation and evolution.

Figure\,\ref{Fig:hist_fast} presents the distributions of key measured parameters --- spectral noise ($\sigma_{\rm rms}$), signal‑to‑noise ratio (SNR), line widths ($W_{50}$, $W_{20}$), heliocentric velocity ($cz_{\odot}$), and \hi mass --- for the FASHI DR2 and ALFALFA $\alpha$100 catalogs. Statistical summaries for the FASHI sample are provided in each panel. The figure demonstrates that FASHI not only detects a significantly larger population of low‑mass galaxies, but also achieves much lower rms noise levels compared to ALFALFA.

\begin{table*}
\caption{\textbf{Cross‑matched catalog of FASHI and ALFALFA sources.}}
\label{tab:cross_alfa}
\vskip 5pt
\centering \small  
\setlength{\tabcolsep}{1.6mm}{
\begin{tabular}{ccccccc|ccccccc}
\hline \hline
\multicolumn{7}{c|}{FASHI}  &  \multicolumn{7}{c}{ALFALFA}   \\ 
\hline
FASHI ID & RA & DEC & c$z_{\odot}$ & $S_\hj$ & $W_{50}$ & log$M_\hj$  &  AGCNr & RA & DEC & c$z_{\odot}$ & $S_\hj$ & $W_{50}$ & log$M_\hj$    \\
   &  deg & deg   &  $\kms$ & mJy & $\kms$ & $M_{\odot}$  &   & deg & deg & $\kms$      & mJy & $\kms$ & $M_{\odot}$ \\
\hline
20260000001&0.003&5.443&12047.9&231.3&1896.9&10.1&105367&0.002&5.443&11983&274&1140&9.9\\ 
20260000002&0.004&24.909&11190.3&295.7&2428.8&10.1&333313&0.004&24.909&11181&313&1800&10.0\\ 
20260000003&0.004&32.702&10592.8&214.9&453.4&9.3&104570&0.007&32.708&10614&245&860&9.6\\ 
20260000004&0.005&15.892&5995.5&173.6&645.2&9.0&331061&0.010&15.872&6007&260&1130&9.3\\ 
20260000008&0.010&23.085&4451.4&164.7&1920.2&9.1&331060&0.010&23.085&4463&160&1960&9.1\\ 
20260000012&0.014&26.016&10419.4&286.8&2572.2&10.0&331405&0.014&26.016&10409&315&2620&10.1\\ 
20260000016&0.028&28.202&16253.8&406.2&1420.4&10.2&102896&0.028&28.202&16254&406&2370&10.5\\ 
20260000028&0.045&1.107&7400.9&208.0&1013.4&9.4&331066&0.048&1.123&7370&214&2300&9.7\\ 
20260000029&0.047&4.275&3853.4&93.0&818.6&8.6&105368&0.045&4.282&3845&83&720&8.7\\ 
20260000036&0.072&27.400&4656.2&115.2&2216.5&9.2&102571&0.072&27.400&4654&104&2000&9.3\\ 
... ...  & & & & & & & ... ... \\
\hline
\end{tabular}}
\begin{flushleft}
\textbf{Notes.} The full catalog containing $\sim$28\,000 matched sources is available in the online supplementary material and at \url{https://fast.bao.ac.cn/cms/article/271/} and \url{https://zcp521.github.io/fashi}. \\
\end{flushleft}
\end{table*}

\begin{figure*}
\centering
\includegraphics[height=0.34\textwidth, angle=0]{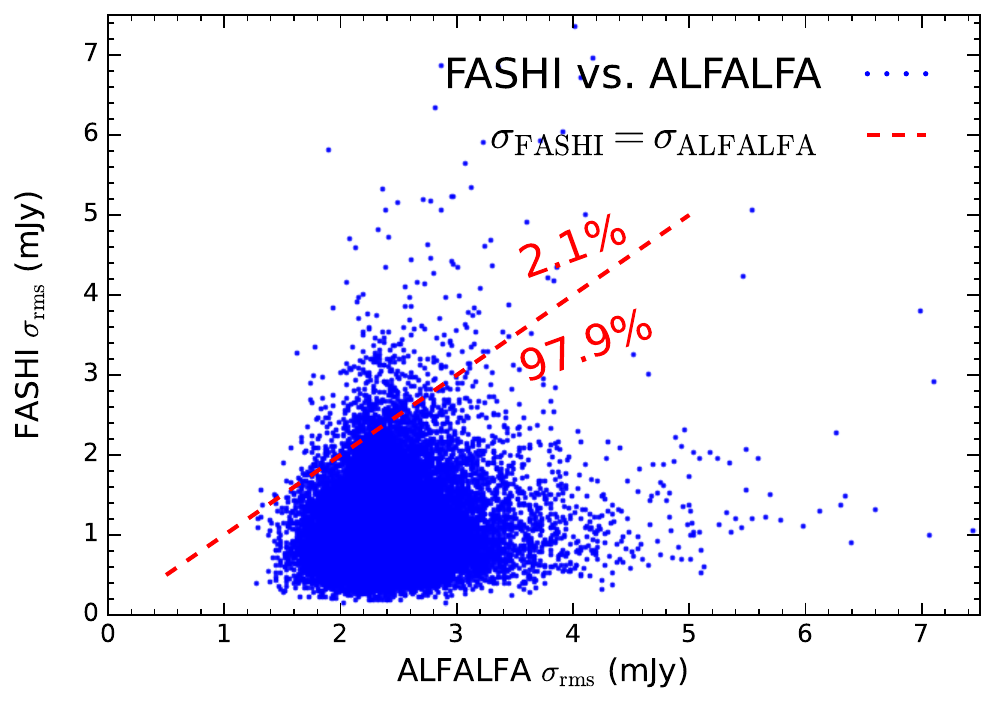}
\includegraphics[height=0.34\textwidth, angle=0]{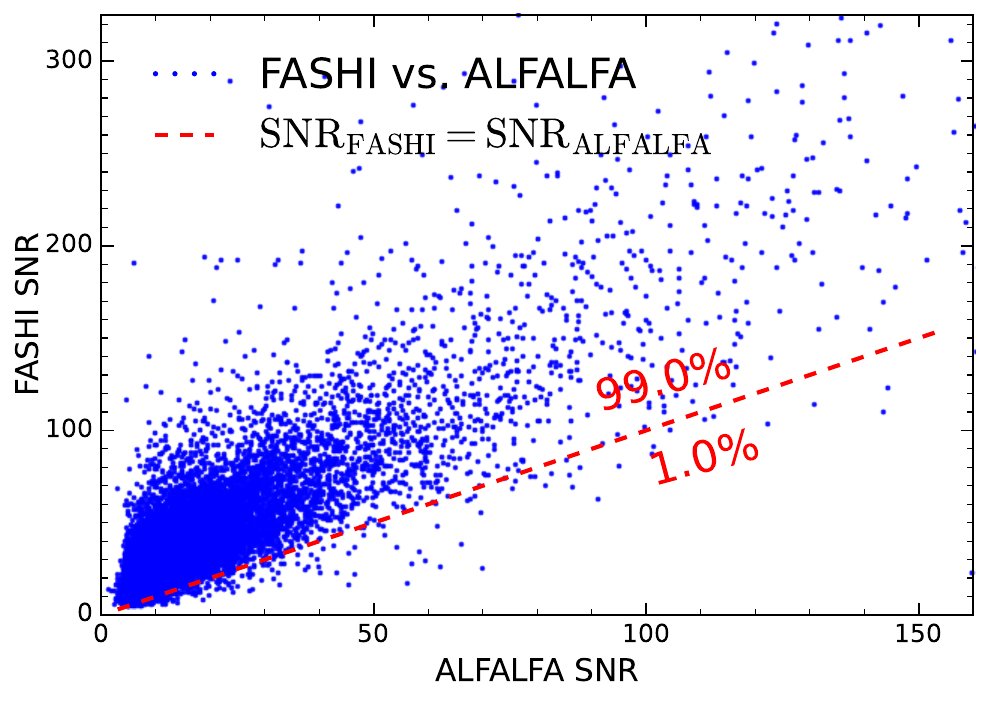}
\includegraphics[height=0.34\textwidth, angle=0]{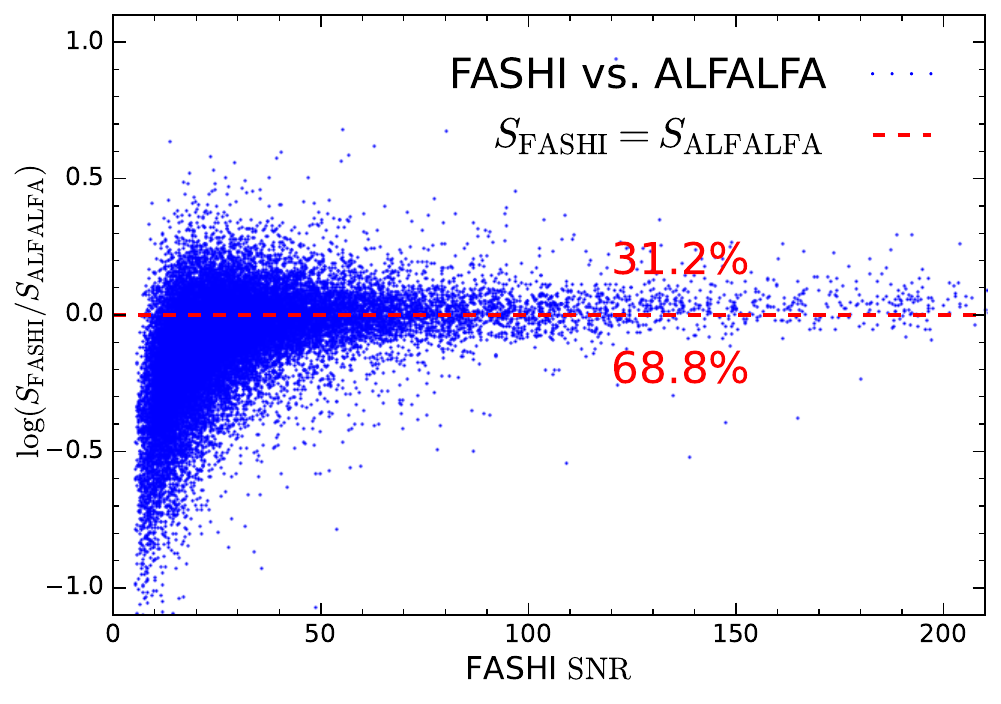}
\includegraphics[height=0.34\textwidth, angle=0]{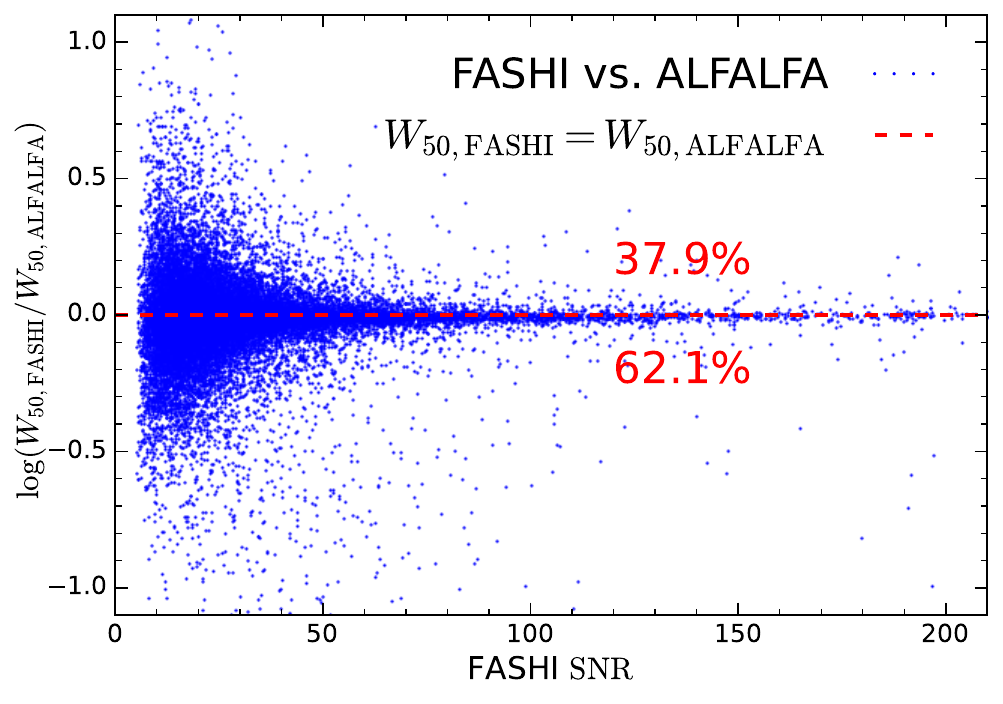}
\includegraphics[height=0.34\textwidth, angle=0]{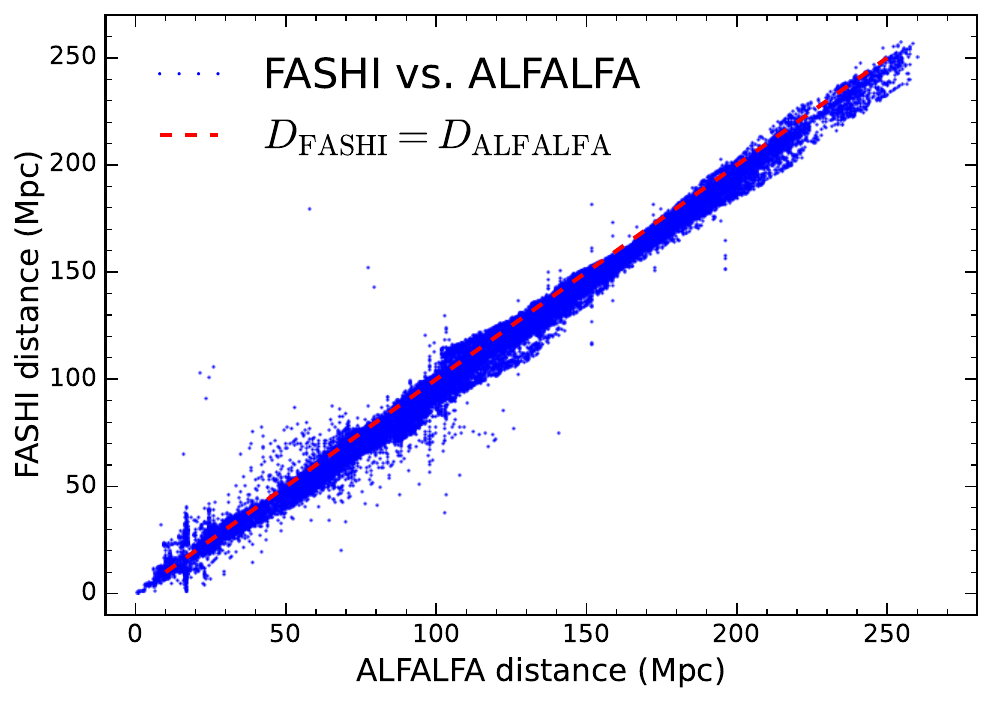}
\includegraphics[height=0.34\textwidth, angle=0]{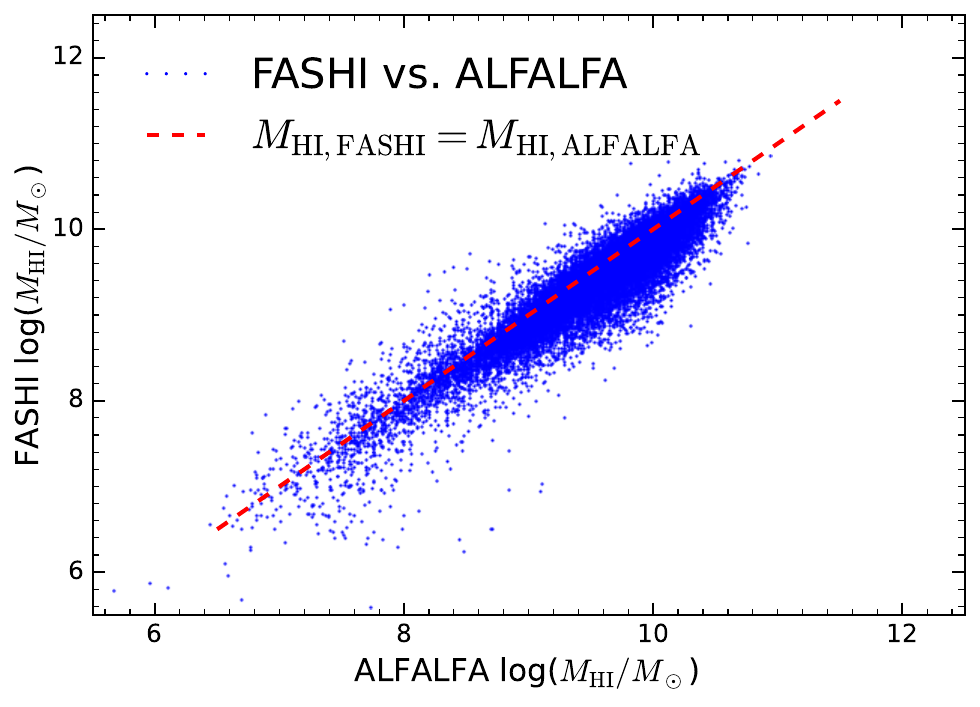}
\caption{FASHI-ALFALFA cross-matched catalog comparison. Comparisons of $\sigma_{\rm rms}$, SNR, flux, $W_{50}$, distance, and mass for over $\sim$28\,000 matched sources (tolerance: $\rm\delta_{RA}\leq 3'$, $\rm\delta_{DEC} \leq 3'$, $\rm\delta_{velocity} \leq 100\,\kms$; see Table~\ref{tab:cross_alfa}$)$. The red dashed line indicates the 1:1 relation. The FASHI $\sigma_{\rm rms}$ is normalized to ALFALFA's $10\,\kms$\ channel width.}
\label{Fig:fast_alfa}
\end{figure*}

\begin{figure*}
\centering
\includegraphics[height=0.205\textwidth, angle=0]{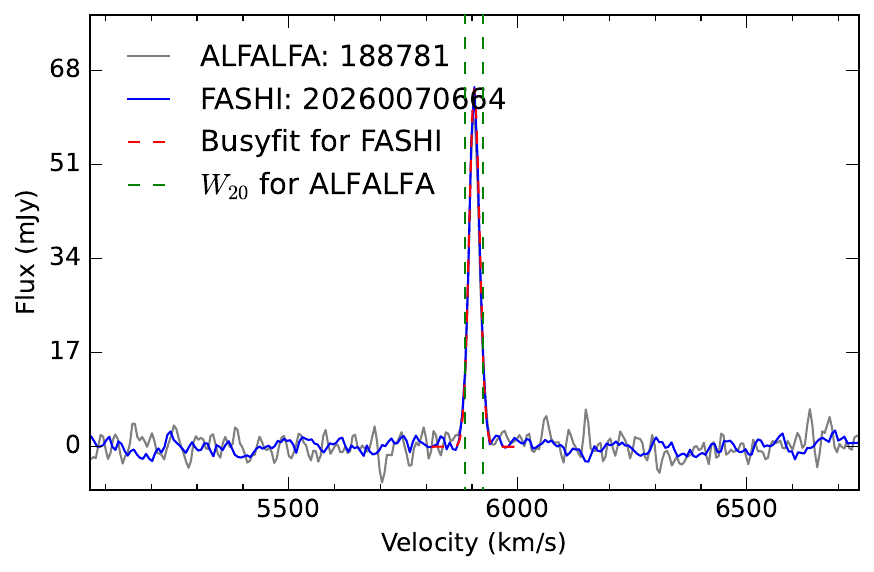}
\includegraphics[height=0.205\textwidth, angle=0]{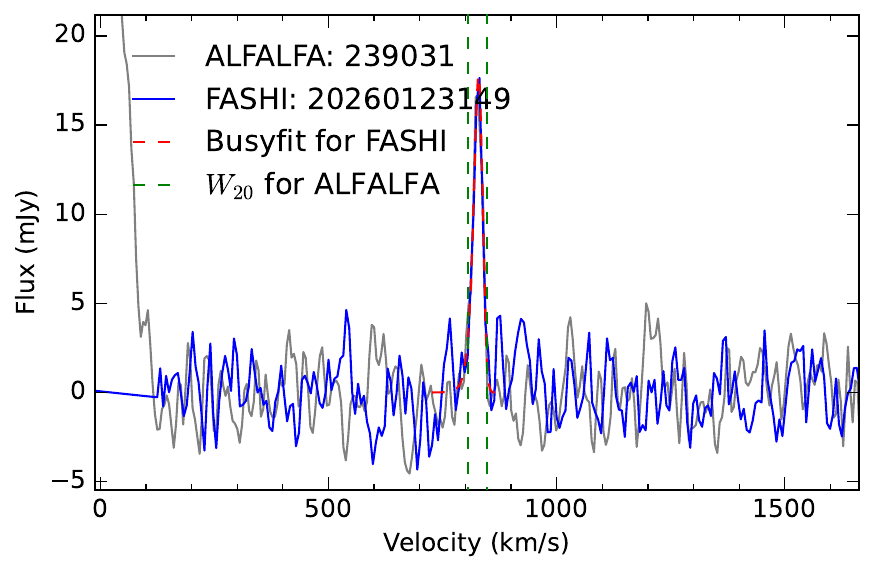}
\includegraphics[height=0.205\textwidth, angle=0]{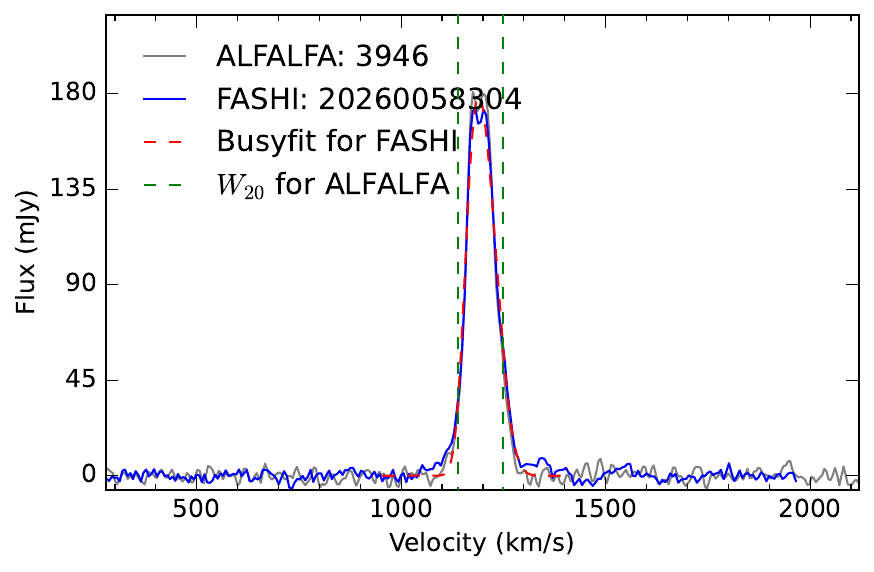}
\includegraphics[height=0.205\textwidth, angle=0]{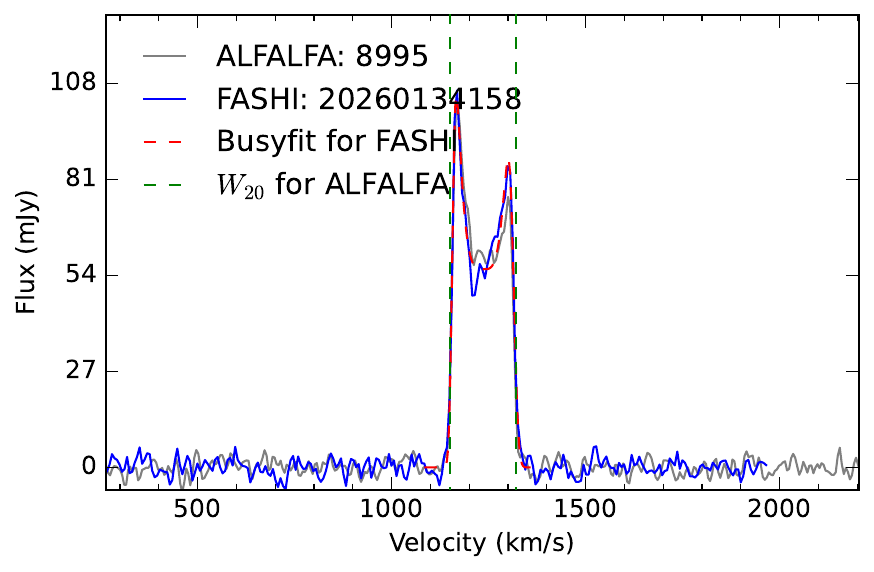}
\includegraphics[height=0.205\textwidth, angle=0]{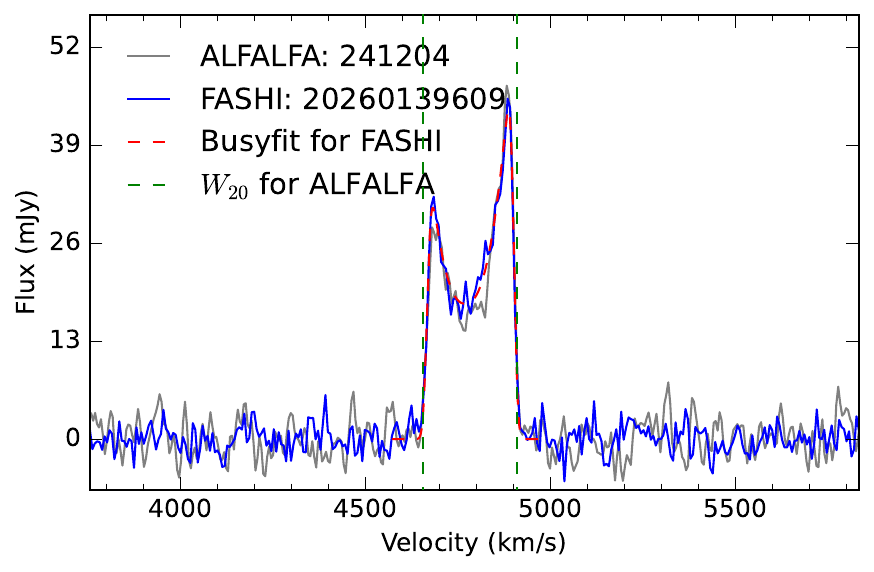}
\includegraphics[height=0.205\textwidth, angle=0]{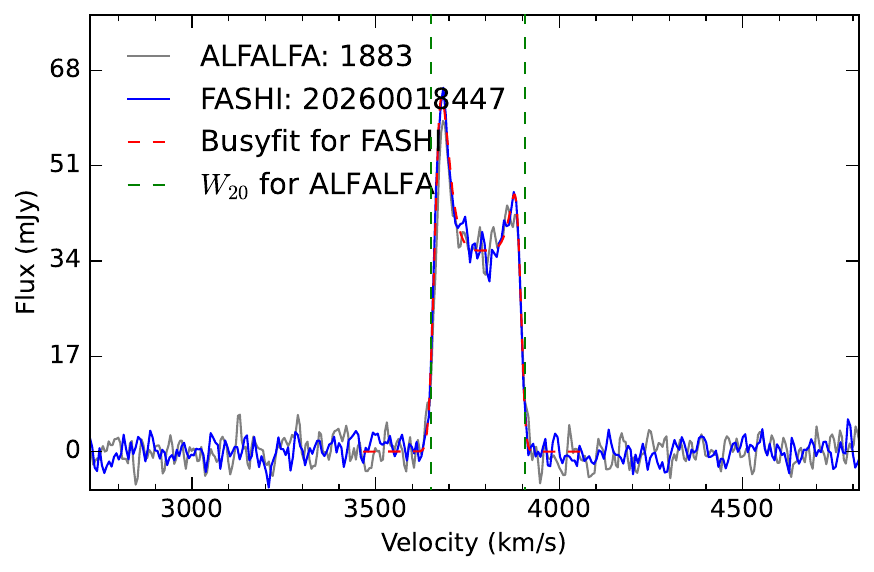}
\includegraphics[height=0.205\textwidth, angle=0]{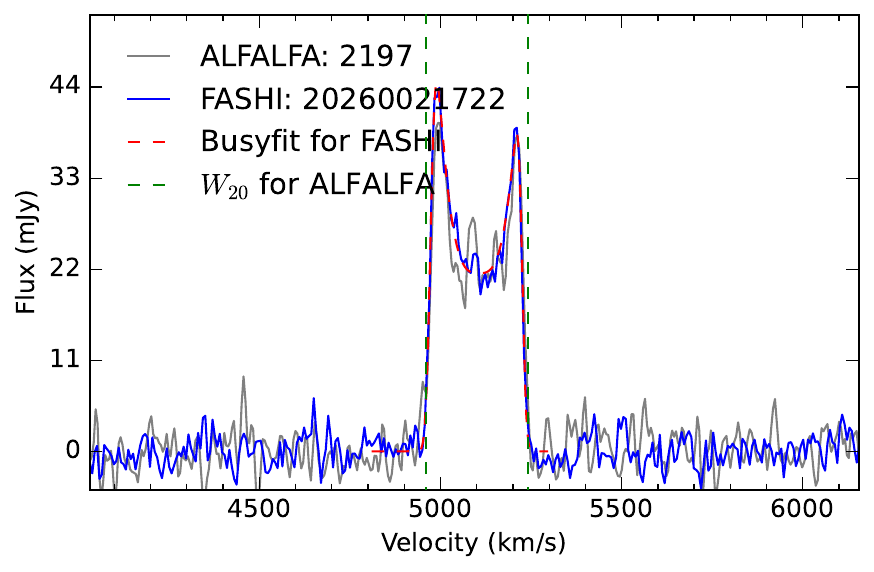}
\includegraphics[height=0.205\textwidth, angle=0]{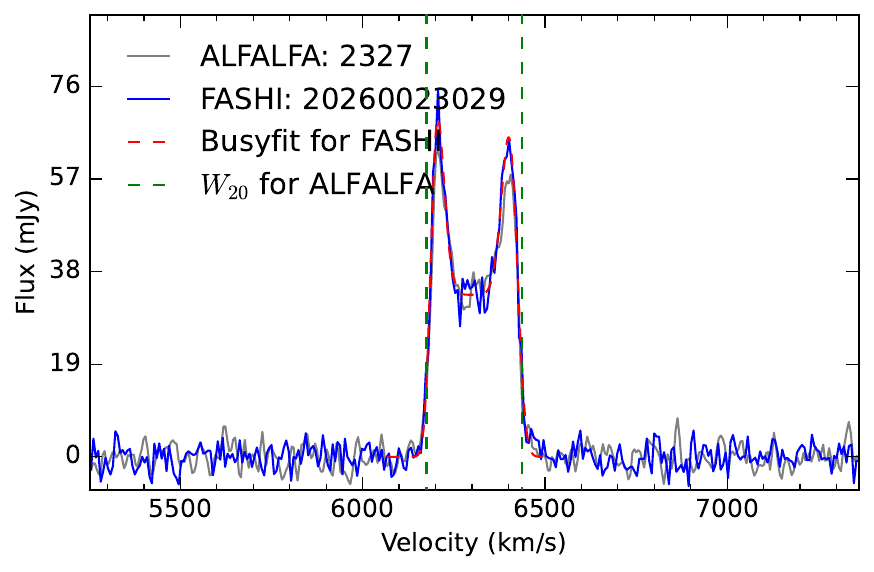}
\includegraphics[height=0.205\textwidth, angle=0]{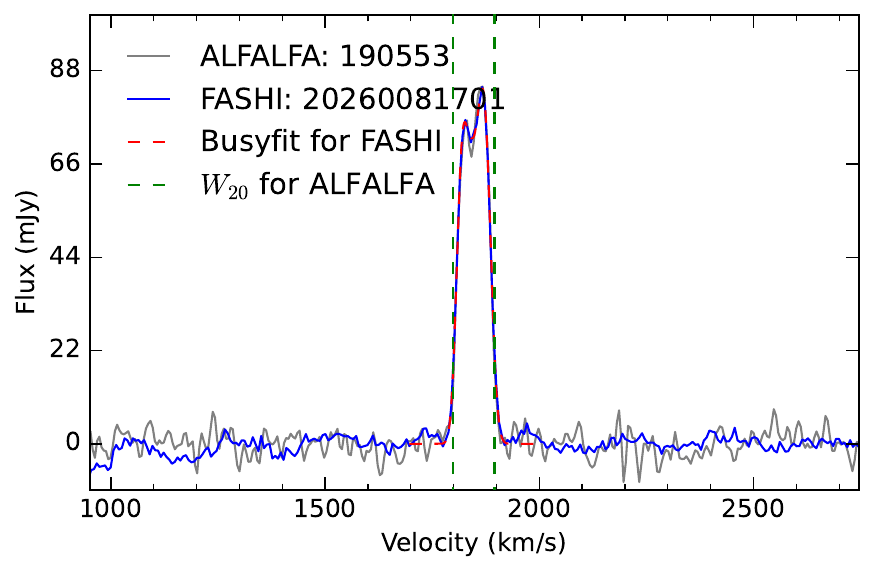}
\includegraphics[height=0.205\textwidth, angle=0]{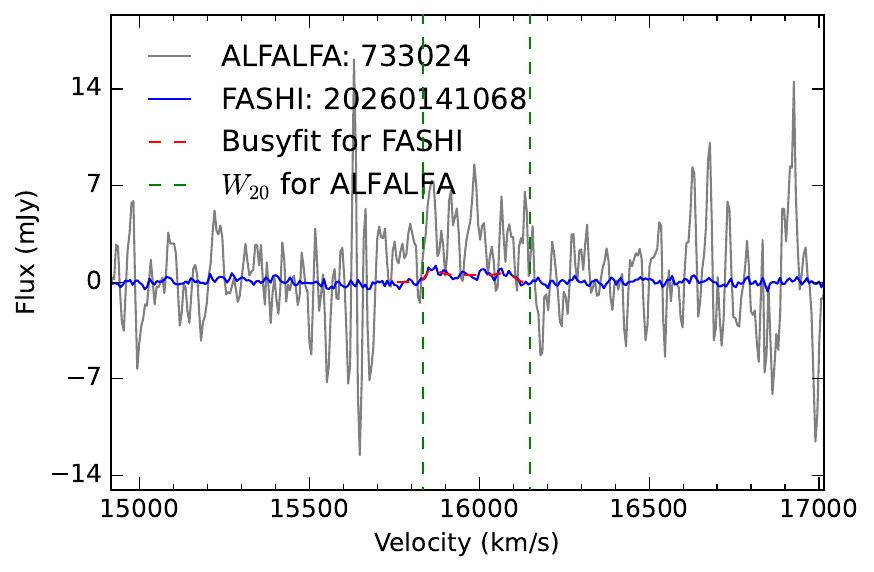}
\includegraphics[height=0.205\textwidth, angle=0]{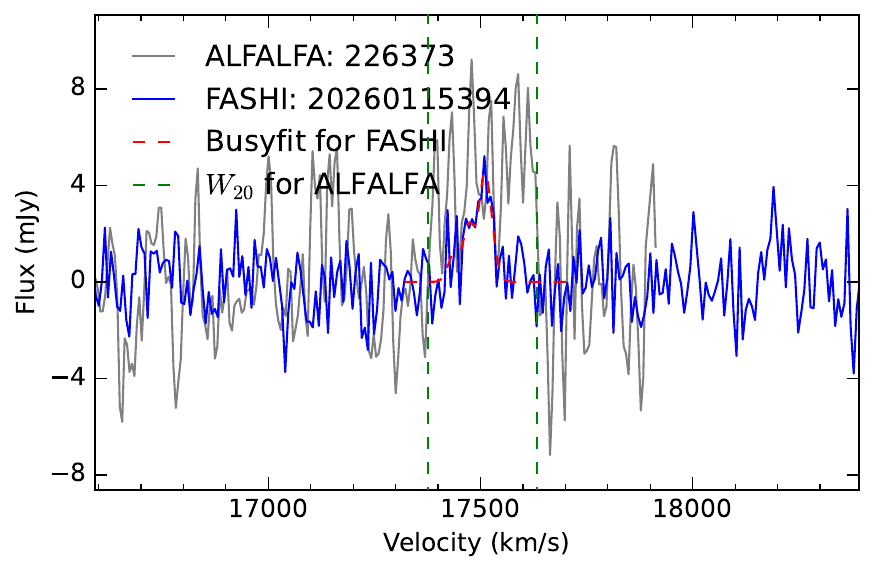}
\includegraphics[height=0.205\textwidth, angle=0]{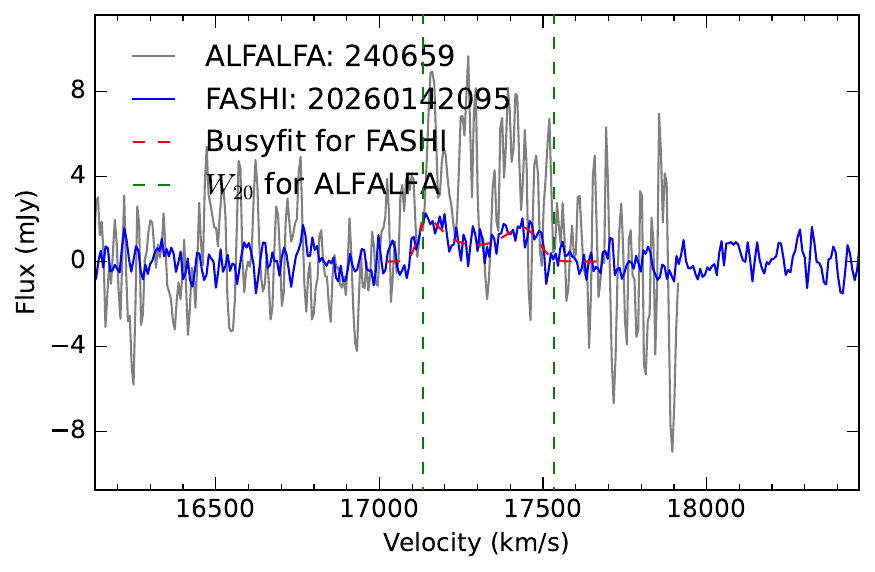}
\includegraphics[height=0.205\textwidth, angle=0]{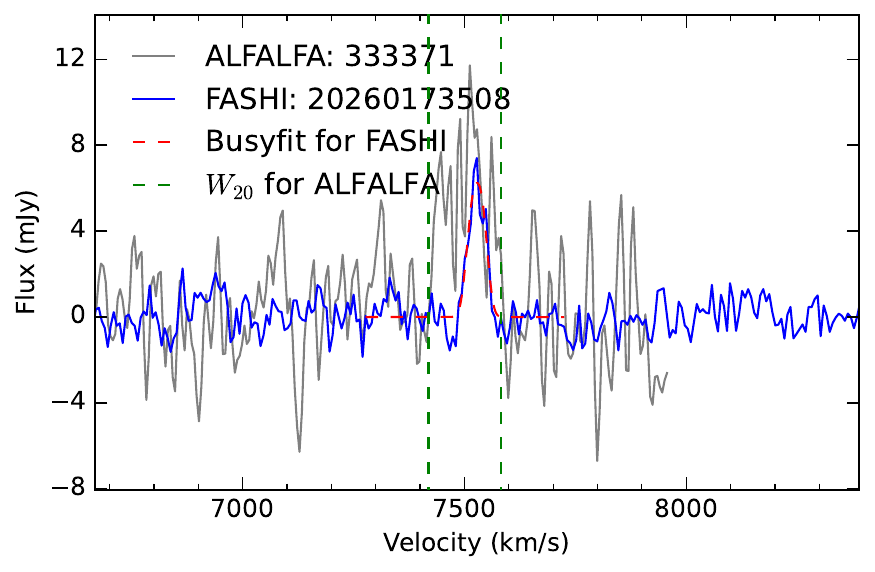}
\includegraphics[height=0.205\textwidth, angle=0]{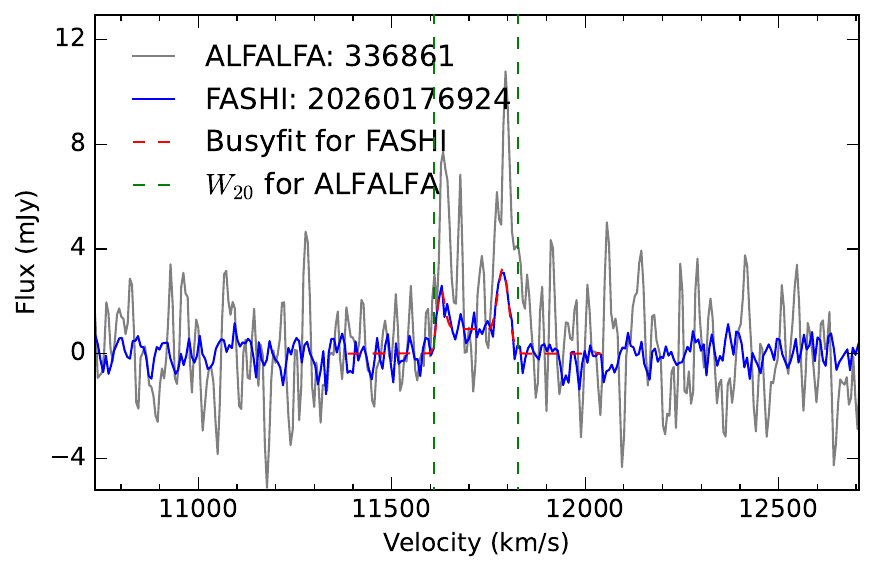}
\includegraphics[height=0.205\textwidth, angle=0]{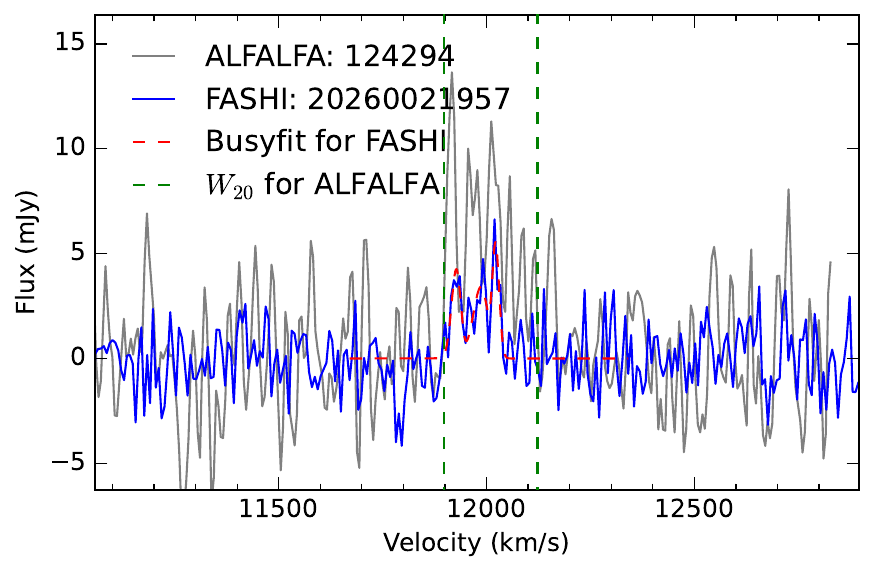}
\includegraphics[height=0.205\textwidth, angle=0]{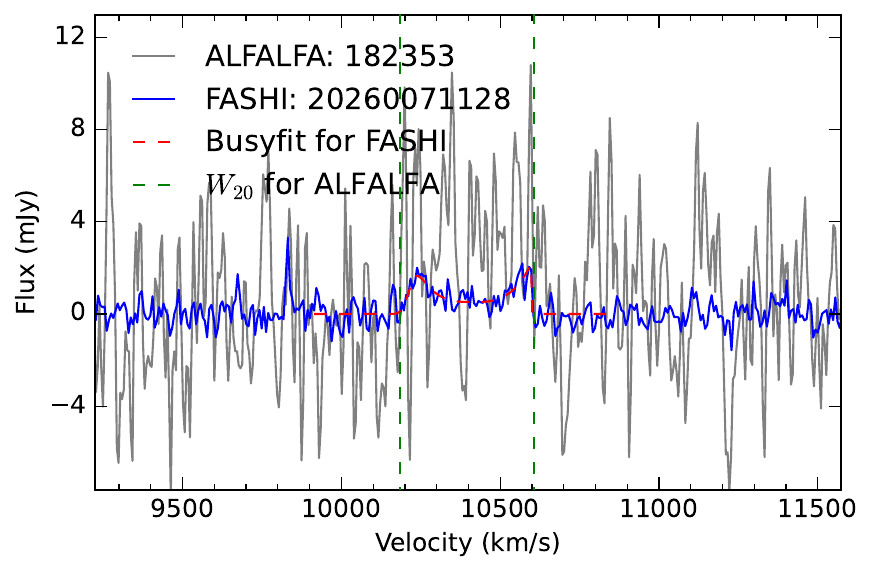}
\includegraphics[height=0.205\textwidth, angle=0]{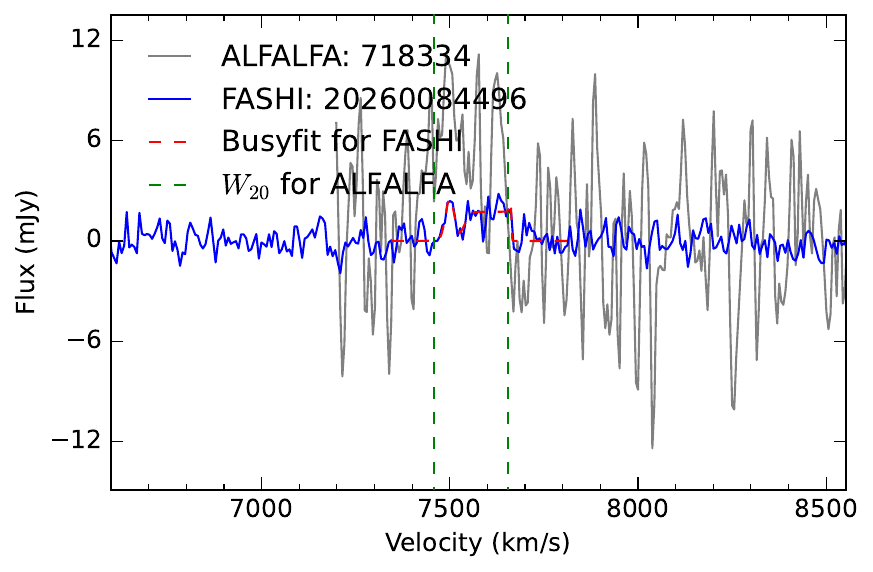}
\includegraphics[height=0.205\textwidth, angle=0]{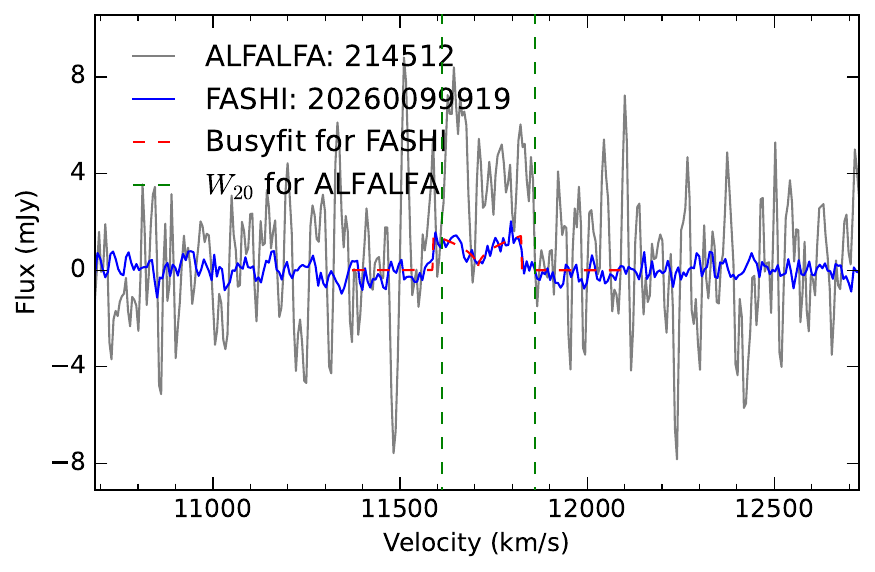}
\caption{Spectra of selected sources comparing FASHI and ALFALFA detections. The upper nine panels show sources with well‑consistent line profiles between the two surveys, while the lower nine panels show sources exhibiting significant flux discrepancies. In each panel, the red curve represents the best-fit function to the FASHI spectrum. Two green vertical lines indicate the line width $W_{20}$ of the ALFALFA spectrum. Both FASHI and ALFALFA source identifiers are also shown.}
\label{Fig:spec_compare}
\end{figure*}

\begin{figure*}
\centering
\includegraphics[height=0.39\textwidth, angle=0]{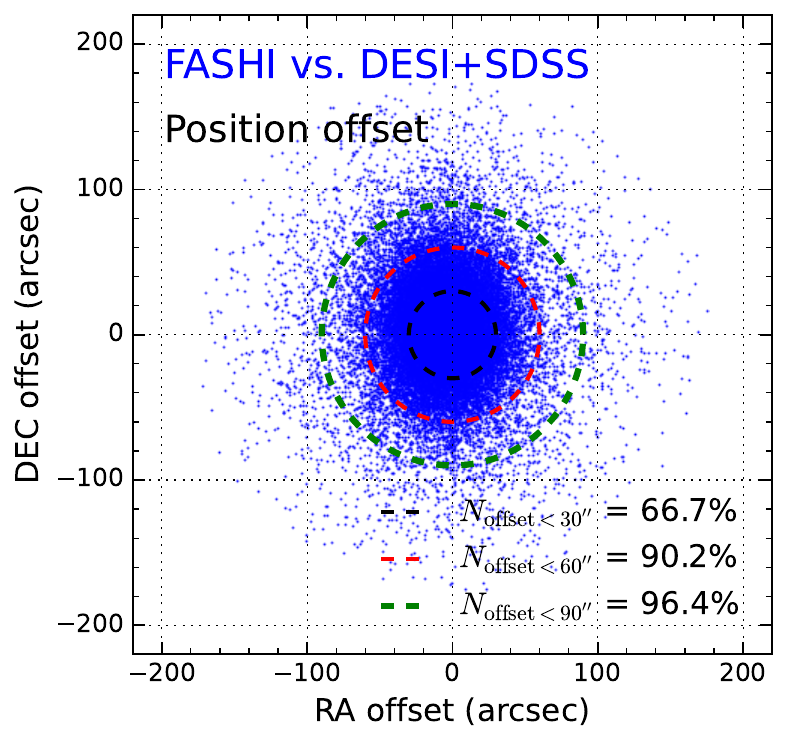}
\includegraphics[height=0.39\textwidth, angle=0]{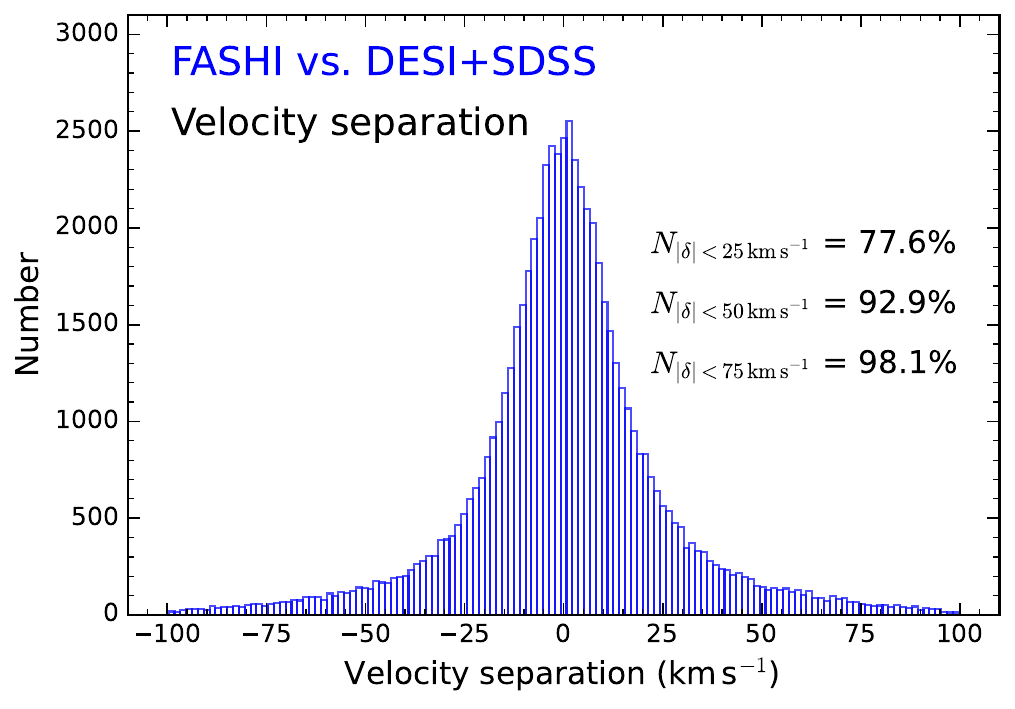}
\caption{Distributions of the RA, DEC, and velocity separations for the 64\,000 FASHI sources matched to DESI and SDSS spectroscopic counterparts (tolerances: $\rm\delta_{RA}\leq 3'$, $\rm\delta_{DEC} \leq 3'$, $\rm\delta_{velocity} \leq 100\,\kms$), displaying the fraction of sources within given offset ranges in each panel.}
\label{Fig:hist_match_sdss}
\end{figure*}

Applying a cross-match between the FASHI and ALFALFA catalogs with positional and velocity tolerances of $\delta_{\rm RA}\leq3'$, $\delta_{\rm DEC}\leq3'$ and $\delta_{\rm velocity}\leq100$\,$\kms$, we obtain $\sim$28\,000 common sources (see Table\,\ref{tab:cross_alfa}; full catalog available online). This well-matched sample enables a detailed comparison of measured parameters, as shown in Figure\,\ref{Fig:fast_alfa}. The comparison reveals that 97.9\% of FASHI sources have lower spectral noise ($\sigma_{\rm rms}$) and 99.0\% have higher signal-to-noise ratios (SNR) than their ALFALFA counterparts when both are compared at a common $10\,\kms$ spectral resolution. Integrated flux densities agree well (see Figure\,\ref{Fig:spec_compare}) for high-SNR sources ($\mathrm{SNR}_{\mathrm{FASHI}} \gtrsim 30$). For faint sources with $\mathrm{SNR}_{\mathrm{FASHI}} \lesssim 20$, however, the FASHI flux measurements are systematically lower than their ALFALFA counterparts. This offset is consistent with Eddington bias affecting the ALFALFA catalog: the higher noise level in ALFALFA preferentially scatters faint sources upward across the detection threshold, leading to systematically overestimated fluxes for low-SNR detections. The superior sensitivity of FASHI minimizes this bias, yielding more reliable flux measurements at the faint end. Direct spectral comparisons for sources with large flux discrepancies (Figure\,\ref{Fig:spec_compare}) further illustrate the superior data quality of FASHI: ALFALFA spectra show significantly higher noise and, in many cases, substantially different measured line widths ($W_{50}$, $W_{20}$). Taken together --- the lower noise, higher SNR, and more consistent line profiles --- we conclude that the FASHI data provide more precise measurements than the ALFALFA survey, especially for faint and low-SNR sources. In addition, the distance and mass parameters in Figure\,\ref{Fig:fast_alfa} exhibit good consistency. This agreement further supports the internal reliability of the released FASHI catalog.

To evaluate the completeness of FASHI relative to ALFALFA, we performed a cross-match between the two catalogs using generous matching criteria ($\delta_{\text{RA}}\leq 6'$, $\delta_{\text{DEC}}\leq 6'$, $\delta_{\text{velocity}}\leq 100\;\text{km}\,\text{s}^{-1}$; these differ from the selection criteria used in Table\,\ref{tab:cross_alfa}). This comparison identifies approximately 2800 ALFALFA sources without counterparts in the current FASHI catalog. These unmatched sources can be attributed to three main factors. First, while FASHI covers approximately 96\% of the ALFALFA footprint, a region of about $300~\deg^{2}$ at $22^{\rm h}<{\rm RA}<23^{\rm h}$ remains unobserved (see Figure\,\ref{Fig:observed_sky}), accounting for 1600 of the unmatched sources. Second, an additional 300 sources fall within areas that are nominally covered but have no usable FASHI data, primarily due to severe RFI. Third, the remaining $\sim$900 ALFALFA sources are located in regions well covered by FASHI where the nominal sensitivity exceeds that of ALFALFA. For these sources, careful inspection of the FASHI data cubes reveals no reliable H\,{\sc i} emission at the reported ALFALFA positions and redshifts. Their absence in the FASHI catalog therefore reflects the higher data quality and more robust source selection criteria adopted in our work, which effectively exclude marginal detections that may have been retained in ALFALFA.

\subsection{Comparison with DESI and SDSS Sources}
\label{sec:match_desi}

The large beam of FAST ($\sim$2$\dotmin$9) compared to optical resolutions ($\sim$1$''$) complicates the identification of unambiguous optical counterparts, particularly in merging systems or for extended \hi features that may not coincide with a single optical galaxy. A detailed study of the galaxy properties will be presented in an upcoming work. To obtain an initial census of the optical properties of the FASHI sample, we cross‑matched the \hi sources with DESI DR1 \citep{desi2025} and SDSS spectroscopic catalogs \citep{Abazajian2009} using positional and velocity tolerances of \(\delta_{\rm RA}\leq3'\), \(\delta_{\rm DEC}\leq3'\) and \(\delta_{\rm velocity}\leq100\)\,$\kms$, yielding $\sim$64\,000 matched sources for further analysis.

Figure\,\ref{Fig:hist_match_sdss} presents the positional and velocity offsets between the FASHI \hi sources and their matched optical counterparts from the DESI DR1 and SDSS spectroscopic catalogs. The cross‑matched sample of $\sim$64\,000 sources shows that 90.2\% have positional offset $<60''$ and 92.9\% have velocity separation $<50\,\kms$. Given the high reliability of the DESI and SDSS positions and redshifts, these small offsets demonstrate that the FASHI catalog provides highly accurate positions and velocities.

\begin{figure}[H]
\centering
\includegraphics[width=0.48\textwidth, angle=0]{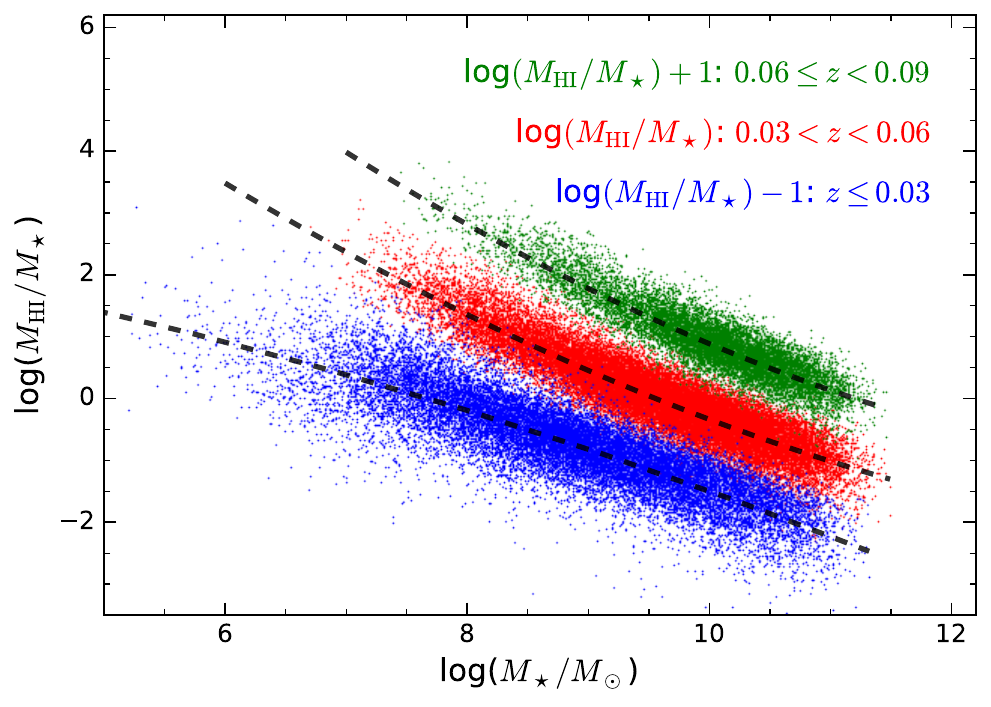}
\caption{The \hi fraction, log($M_\hj/M_\star$), as a function of stellar mass, $\log(M_\star)$, for the FASHI sample in different redshift bins. Data points represent optical counterparts identified by cross‑matching FASHI \hi sources with spectroscopic galaxies from DESI and SDSS. Dashed lines show the median relations for each redshift bin.}
\label{Fig:frac_mass}
\end{figure}

In Figure\,\ref{Fig:frac_mass}, we show the relation of the \hi fraction, $\log(M_\hj/M_\star)$ as a function of stellar mass, $\log M_\star$, for the matched FASHI-DESI sample across different redshift bins, represented by points of different colors. There is no strong evolution of $\log(M_\hj/M_\star)$ in the three redshift bins. We separate the measurements by an offset of 1~dex for better comparisons. The general trend of $\log(M_\hj/M_\star)$ is consistent with previous studies \citep{Huang2012}. Detailed comparisons will be presented in the future work. We note that at the low-mass end, the slope of $\log(M_\hj/M_\star)$ becomes substantially steeper for higher redshift bins. It is caused by the flux-limited sample selection effects of both FASHI and DESI surveys. For low-mass galaxies, only those luminous and gas-rich ones would fall within the flux limits.

\section{Discussion}\label{sec:discuss}

\subsection{Reliability}
The FASHI catalog (Table~\ref{tab:exgalcat}) was constructed by extracting sources from the data cubes using \texttt{SoFiA} with a $4.5\sigma$ detection threshold. All candidates were subsequently subjected to visual inspection based on moment maps (0$^{\rm th}$, 1$^{\rm st}$, and 2$^{\rm nd}$), integrated spectra, and signal-to-noise ratios (SNR $>4$). This manual verification effectively removed spurious detections caused by residual radio-frequency interference, baseline instabilities, or edge effects, and refined source boundaries. In the final catalog, approximately 64\,000 sources have secure spectroscopic counterparts in DESI and SDSS, with positional offsets smaller than $60''$ and velocity differences within $50~\kms$, demonstrating the high astrometric and spectral fidelity of the FASHI measurements. Comparisons with ALFALFA further show that FASHI extends to fainter integrated fluxes and lower \hi masses, consistent with its improved sensitivity.

Independent validation of the FASHI catalog has been provided by follow-up observations with high-resolution interferometers and single-dish telescopes. Previous studies have reported excellent agreement between FASHI measurements and observations from the Very Large Array (VLA), the Green Bank Telescope (GBT), and the Arecibo telescope \citep[e.g.,][]{Benitez2024, Karunakaran2024, Zhang2025, Siljeg2026}. For example, a detailed comparison of source J0139+4328 by \citet{Siljeg2026} shows that the integrated \hi flux, peak flux density, and systemic velocity derived from FASHI are consistent with the VLA results within the quoted uncertainties, with velocity differences typically $\lesssim1~\kms$, well below the FASHI channel width of $6.4~\kms$. These independent consistency checks confirm the robustness of the FASHI data reduction pipeline and support the reliability of the catalog for statistical studies of \hi properties in the local Universe.

\subsection{Sample Completeness}
\label{sec:completeness}

\begin{figure}[H]
\centering
\includegraphics[width=0.48\textwidth]{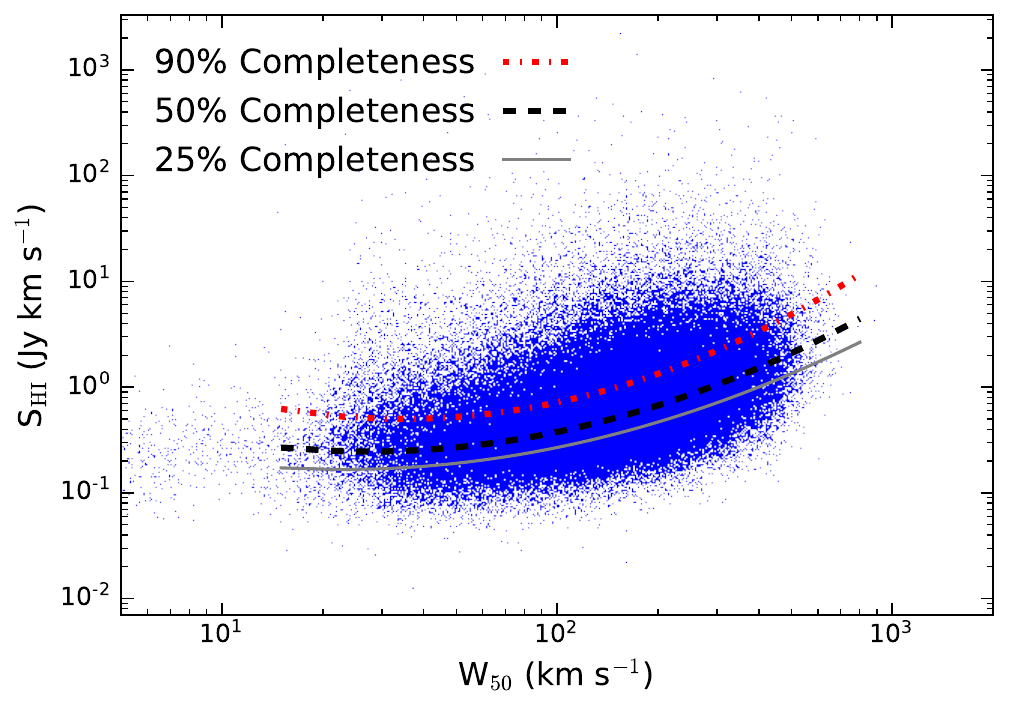}
\caption{The integrated \hi flux $\log S_\hj$ versus line width $\log W_{50}$ for the FASHI sample. The 25\%, 50\%, and 90\% completeness limits for $f_\sigma=0.649$ are indicated by different curves.}
\label{Fig:completeness}
\end{figure}

Accurate completeness estimation is essential for measuring the H\,{\sc i} mass function (HIMF). \citet{Haynes2011} established a 50\% completeness limit for the integrated 21~cm flux density in the ALFALFA sample and restricted their analysis to galaxies above this threshold. A similar approach was adopted for FASHI DR1 by \citet{Ma2025}. However, FASHI faces an additional challenge: the non-uniform distribution of detection sensitivity. As shown in Figure\,\ref{Fig:sensitivity}, the per-source sensitivity $f_\sigma$ varies significantly across the survey footprint due to the schedule-filler observing strategy, resulting in spatially varying completeness limits.

To account for this, \citet{Ma2025} applied correction weights to balance the variation in surface number density across different sky pixels, achieving nearly homogeneous weighted surface density. In FASHI DR1, weights were derived separately for the northern ($30\degree<{\rm DEC}<66\degree$) and southern ($-6\dotdeg2<{\rm DEC}<0\degree$) strips \citep{Ma2025}, with the 50\% completeness limit in the north found to be 1.5~dex deeper than in the south. However, applying similar weighting schemes to the full FASHI DR2 sample would yield an average 50\% completeness limit that is overly optimistic for low-$f_\sigma$ (high surface density) regions, placing an excessive number of \hi detections below the nominal completeness threshold.

To better exploit the statistical power of the FASHI DR2 sample, we develop an improved method that explicitly accounts for the inhomogeneous sky coverage. In principle, the varying survey depth introduces an additional dependence of the flux limit on the detection sensitivity; that is, we model the completeness of each galaxy as $C(S_\hj|W_{50},f_\sigma)$. Following \citet{Ma2025}, we fit the measurements of $S_\hj^{3/2}\,\mathrm{d}n/\mathrm{d}\log S_\hj$ in each $W_{50}$ and $f_\sigma$ bin with an error function:
\begin{equation}
C(S_\hj|W_{50},f_\sigma)=\frac{1}{2}\left[1+\mathrm{erf}\left(\frac{\log S_\hj-\log S_{\hj,50\%}}{\sigma_{\log S_\hj}}\right)\right],\label{eq:comp}
\end{equation}
where $S_{\hj,50\%}$ is the 50\% completeness limit and $\sigma_{\log S_\hj}$ characterizes the steepness of the completeness decline at low fluxes. An additional free parameter is used to fit the amplitude of $S_\hj^{3/2}\,\mathrm{d}n/\mathrm{d}\log S_\hj$. 

\begin{figure}[H]
\centering
\includegraphics[width=0.48\textwidth]{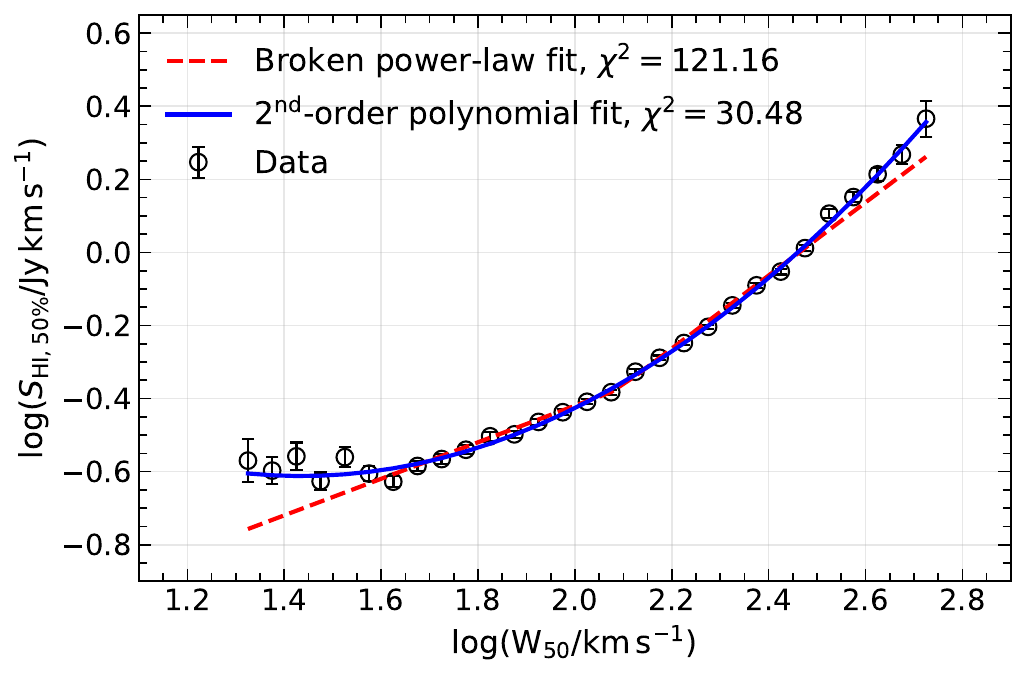}
\caption{The best-fitting $\log S_{\hj,50\%}$ values in different $\log W_{50}$ bins. The blue and magenta lines show the best-fitting second-order polynomial and broken power-law functions to fit the dependence of $\log S_{\hj,50\%}$ on $W_{50}$, respectively.}
\label{fig:w50s21}
\end{figure}
\begin{figure*}
\centering
\includegraphics[width=\textwidth]{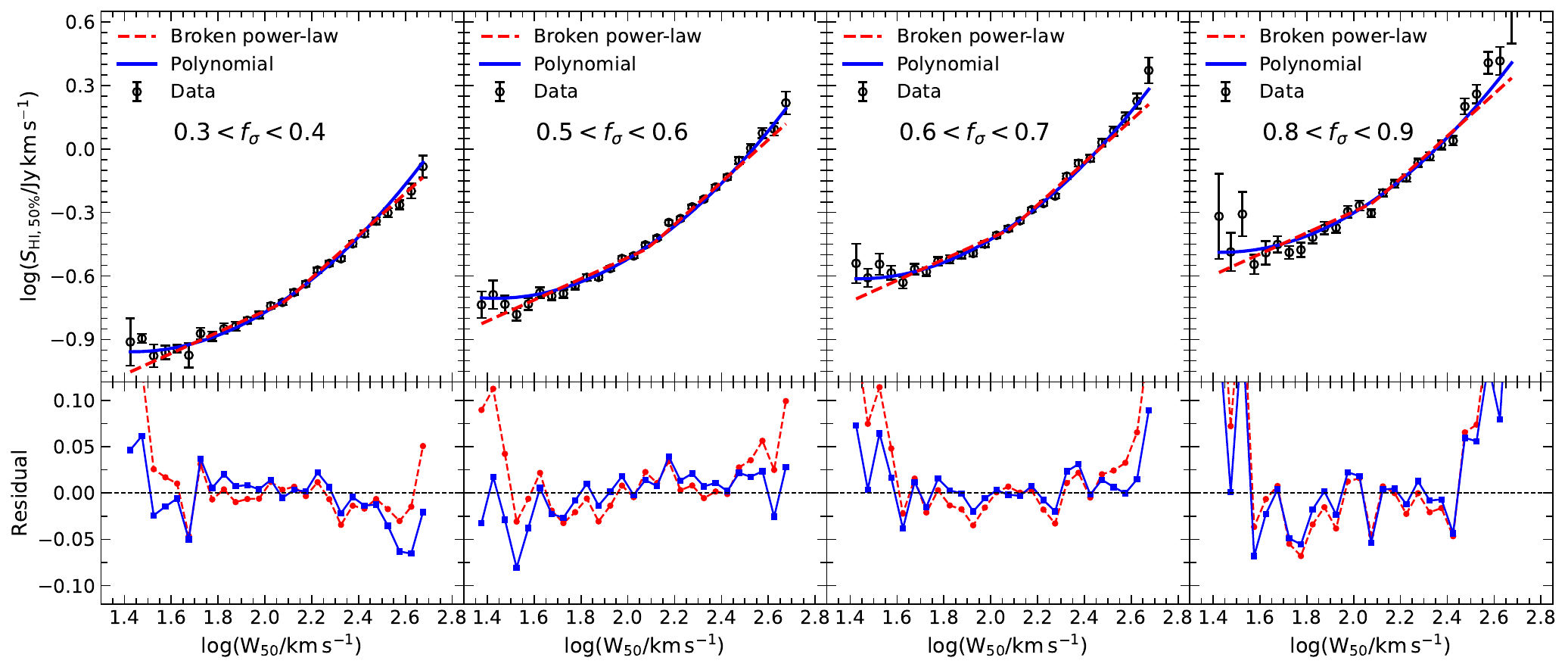}
\caption{Top: Similar to Figure~\ref{fig:w50s21}, but for the samples in different $f_\sigma$ bins. Bottom: The differences between the model fittings and the measured $S_{\hj,50\%}$ values, with red lines for broken power-law fits and blue lines for second-order polynomial fits. }
\label{fig:w50s21_fsigma}
\end{figure*}
With substantially improved sampling of galaxies in low-$W_{50}$ bins, we find that the dependence of $S_{\hj,50\%}$ on $W_{50}$ is better described by a second-order polynomial than by the broken power-law adopted in previous studies \citep{Haynes2011,Oman2022,Ma2025}. The broken power-law significantly underestimates $S_{\hj,50\%}$ for $W_{50}<30\,\kms$, a regime not well probed by earlier surveys. As shown in Figure~\ref{fig:w50s21}, the $\chi^2$ value of the second-order polynomial fit is smaller than that of broken power-law by $\Delta\chi^2=90.67$. It leads to an Akaike Information Criterion (AIC) difference of $\Delta{\rm AIC}=80.67$ or a Bayesian Information Criterion (BIC) difference of $\Delta{\rm BIC}=87.31$, which strongly favors the second-polynomial fits for our measurements. By further splitting each $W_{50}$ bin into $f_\sigma$ bins as in Figure~\ref{fig:w50s21_fsigma}, we find that the additional dependence on $f_\sigma$ follows a double power-law relation. The best fit is given by:
\begin{flalign}
    &\log S_{\hj, 50\%} = 0.586 - 1.670 \log W_{50} + 0.582 (\log W_{50})^{2}+ \nonumber\\
    &\quad f_\sigma-0.362-\log\left[10^{0.775(f_\sigma-0.447)}+10^{-1.485(f_\sigma-0.447)}\right].\label{eq:shi50}
\end{flalign}

The scatter $\sigma_{\log S_\hj}$ is also better represented by a second-order polynomial:
\begin{equation}
\sigma_{\log S_\hj} = 0.935 - 0.649 \log W_{50} + 0.167 (\log W_{50})^{2}.
\end{equation}
The dependence of $\sigma_{\log S_\hj}$ on $f_\sigma$ is negligible, so we model it as a function of $W_{50}$ only. For galaxies with low $f_\sigma$, the 50\% completeness limit is substantially lower. At fixed $W_{50}$, $S_{\hj,50\%}$ for $f_\sigma = 0.2$ is 0.63~dex smaller than for $f_\sigma = 0.649$ (where the $f_\sigma$ correction term in Eq.\,(\ref{eq:shi50}) vanishes).

Following common practice, we also derive the 25\% and 90\% completeness limits. These are computed analytically from Eq.\,(\ref{eq:comp}) as:
\begin{align}
\log S_{\hj,25\%} &= \log S_{\hj,50\%} - 0.477\,\sigma_{\log S_\hj}, \\
\log S_{\hj,90\%} &= \log S_{\hj,50\%} + 0.906\,\sigma_{\log S_\hj}.
\end{align}
Figure\,\ref{Fig:completeness} shows the 25\%, 50\%, and 90\% completeness curves for $f_\sigma = 0.649$ as solid, dashed, and dotted lines, respectively. The derived completeness values are provided in Table\,\ref{tab:exgalcat}.

\subsection{\hi mass function}
\label{sec:HIMF}

\begin{figure*}
\centering
\includegraphics[width=0.49\textwidth]{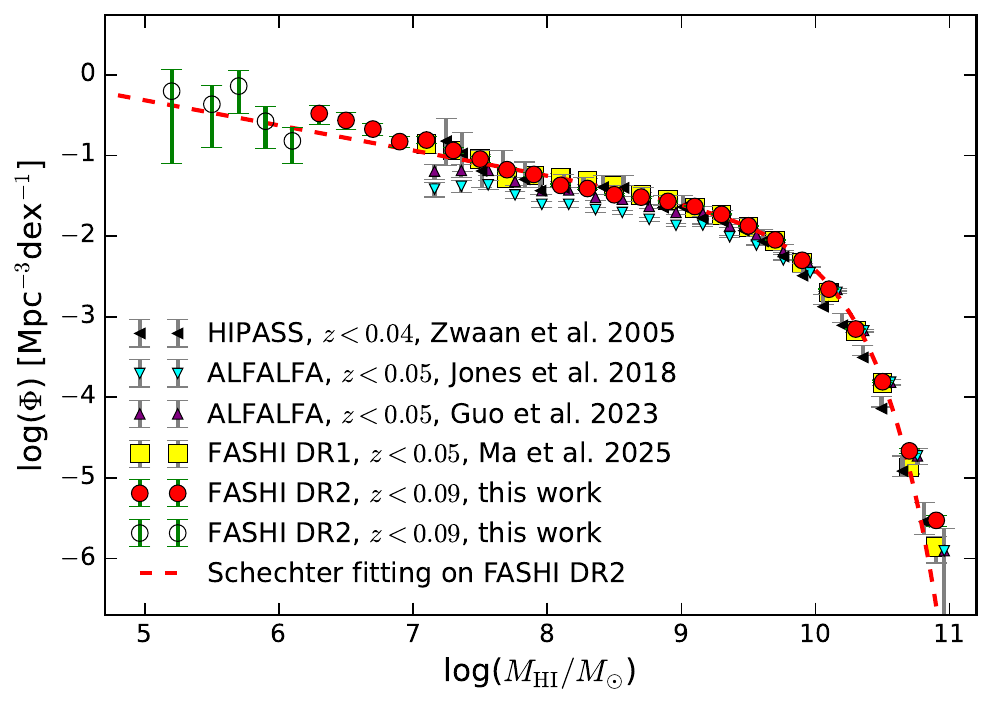}
\includegraphics[width=0.49\textwidth]{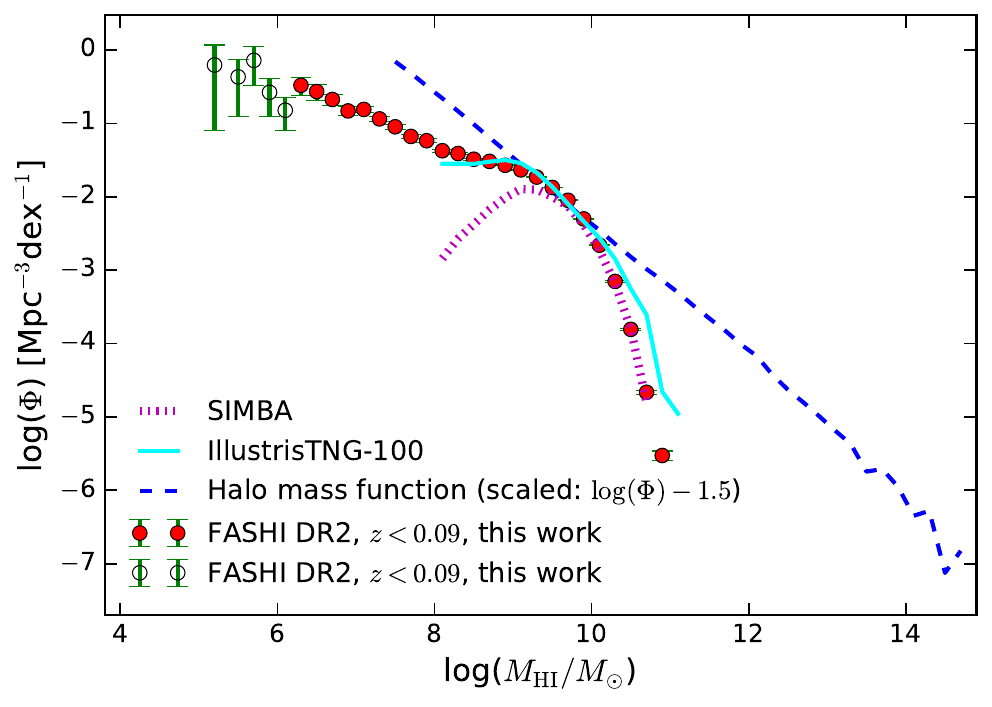}
\caption{Left: The HIMF derived from the FASHI DR2 sample (filled circles), in comparisons to previous measurement from HIPASS \citep{Zwaan2005}, ALFALFA \citep{Jones2018,Guo2023} and FASHI DR1 \citep{Ma2025}. Considering the systematic uncertainties, our HIMF measurements are more robustly estimated at $M_\hj>10^{6.2}\msun$. The measurements below this mass is shown for completeness (open circles). The best-fitting Schechter function is displayed as the red dashed line. Right: Our HIMF in comparisons with the simulation predictions of IllustrisTNG-100 \citep{Diemer2018} and SIMBA \citep{Dave2020}. The halo mass function from IllustrisTNG is shown as the blue dashed line, scaled to match the amplitude of HIMF. }
\label{Fig:himf}
\end{figure*}

With accurate completeness estimates in hand, we derive the \hi mass function (HIMF) following \citet{Ma2025}:
\begin{equation}
    \phi(M_{\hj}) = \sum_i \frac{1}{C(S_{\hj,i} | W_{50,i}, f_{\sigma,i}) \, V_{\mathrm{max},i}},
\end{equation}
where $V_{\mathrm{max},i}$ is the maximum comoving volume (listed in Table\,\ref{tab:exgalcat}) within which the $i$-th galaxy could be detected. To construct a clean sample for HIMF measurement, we apply the following selection criteria:
\begin{align}
    S_\hj &> S_{\hj,50\%}, \\
    W_{20} &< 5 W_{50}, \\
    f_\sigma &< 2, \\
    0 &< z < 0.09.
\end{align}
As in previous studies, we include only galaxies above the 50\% flux completeness limit $S_{\hj,50\%}$. The $V_{\mathrm{max},i}$ for each source is computed directly from Eq.\,(\ref{eq:himass}). The cuts on $W_{20}$ and $f_\sigma$ remove galaxies with unreasonably narrow line widths or excessively high noise levels. After applying all cuts, we retain over $109\,000$ galaxies, corresponding to $70\%$ of the full sample --- substantially larger than any previous sample used for HIMF determination.

The resulting HIMF is shown as filled circles in the left panel of Figure~\ref{Fig:himf}, with the measured data listed in Table~\ref{tab:himf}. For comparison, we also plot the measurement from \citet{Ma2025}, which combined HIPASS, ALFALFA, and FASHI DR1. The HIMF measurements from HIPASS \citep{Zwaan2005} and ALFALFA \citep{Jones2018,Guo2023} are also shown for comparison. We note that \cite{Guo2023} used the 90\% completeness limit to correct for the incompleteness of low-mass galaxies in \cite{Jones2018}. In general, our results are in agreement with all these previous measurements, but we robustly extend the HIMF down to $M_\hj = 10^{6.2}\,\msun$. The measurements below $M_\hj < 10^{6.2}\,\msun$ (open circles) are shown for completeness but are not used to draw conclusions on the global low-mass slope, because all such sources lie within 5 Mpc and are likely affected by Local Volume structure. The range $10^{6.2}\,\msun < M_\hj < 10^7\,\msun$, which extends to $\sim20\,$Mpc provides the more robust low-mass constraint, with estimated systematic uncertainties below $70\%$, as will be shown in the following section.

In the right panel of Figure~\ref{Fig:himf}, we compare our HIMF measurements with the simulation predictions of IllustrisTNG \citep{Diemer2019} and SIMBA \citep{Dave2020}. Remarkably, the low-mass slope shows no evidence of strong steepening toward the lowest masses. The halo mass function combining the measurements of IllustrisTNG-50 \citep{Pillepich2019} and IllustrisTNG-100 \citep{Pillepich2018} is shown as the blue dashed line, scaled to match the amplitude of our HIMF. Over the robustly constrained range $M_\hj>10^{6.2}\msun$, the HIMF remains substantially shallower than a scaled halo mass function. This comparison suggests that the occupation of detectable \hi reservoirs in low-mass halos is highly inefficient. However, because the lowest-mass bins are affected by Local Volume structure and large systematic uncertainties, we do not use the present data to infer the detailed \hi occupation of halos below $10^{6.2}\msun$.

We also fit the HIMF measurements with the conventional Schechter function,
\begin{equation}
\phi(M_{\hj}) =
\ln10\,\phi_s\left( \frac{M_\hj}{M_s} \right)^{\alpha+1}\exp\left( -\frac{M_\hj}{M_s} \right),
\label{eq:single_schechter_himf}
\end{equation}
where $\phi_s$, $M_s$ and $\alpha$ are the normalization, characteristic \hi mass and low-mass slope, respectively. Because the lowest-mass bins in Figure~\ref{Fig:himf} are affected by large measurement uncertainties and small-number statistics, we perform the fiducial parametric fit over the range $\log(M_\hj/\msun)>6.2$. To robustly quantify the fitting parameters, we consider total uncertainties by adding the statistical errors and systematic uncertainties in quadrature (see the following section). The best-fitting single-Schechter parameters are
$\phi_s =(6.38 \pm 0.49)\times 10^{-3}h_{70}^{3}{\rm Mpc}^{-3}{\rm dex}^{-1}$,
$\log(M_s/h_{70}^{-2}\msun)=9.89 \pm 0.02$, and $\alpha = -1.31 \pm 0.02$, which is in good agreement with the FASHI DR1 fittings of \cite{Ma2025}.

We also test whether the additional flexibility of a double-Schechter function is statistically warranted. Using only the statistical errors, the double-Schechter form gives a formally better fit, with $\Delta{\rm AIC}={\rm AIC}_{\rm single}-{\rm AIC}_{\rm double}=30.93$ and $\Delta{\rm BIC}={\rm BIC}_{\rm single}-{\rm BIC}_{\rm double}=27.40$. However, this preference is driven mainly by the very small statistical errors in the low-mass bins. When the systematic uncertainty is taken into account, the single-Schechter model is even slightly preferred, with $\Delta{\rm AIC}=-0.33$ and $\Delta{\rm BIC}=-3.86$. 

We therefore adopt the single-Schechter function as the fiducial parametric description of the FASHI HIMF. The double-Schechter model remains a useful phenomenological description of possible curvature at the low-mass end, but the current measurements do not support a robust second Schechter component. Importantly, the integrated cosmic \hi density is insensitive to this choice, because the lowest-mass galaxies contribute only a minor fraction of $\Omega_\hj$.

\begin{table}[H]
\caption{\textbf{HIMF measurement of FASHI.}}
\label{tab:himf}
\vskip 5pt
\centering \small  
\setstretch{0.9}
\setlength{\tabcolsep}{2.0mm}{
\begin{tabular}{cccc}
\hline \hline
$\log M_\hj$ &  $\phi(M_\hj)$ & ${\rm stat. err}_{\phi(M_\hj)}$ & ${\rm sys. err}_{\phi(M_\hj)}$ \\
 $\msun$ & $\rm Mpc^{-3}dex^{-1}$ & $\rm Mpc^{-3}dex^{-1}$ & $\rm Mpc^{-3}dex^{-1}$ \\
\hline
5.2 & $6.249 \times 10^{-1}$ & $5.456 \times 10^{-1}$ & $8.171 \times 10^{-1}$ \\
5.5 & $4.304 \times 10^{-1}$ & $3.045 \times 10^{-1}$ & $7.172 \times 10^{-1}$ \\
5.7 & $7.230 \times 10^{-1}$ & $3.920 \times 10^{-1}$ & $5.426 \times 10^{-1}$ \\
5.9 & $2.650 \times 10^{-1}$ & $1.425 \times 10^{-1}$ & $3.021 \times 10^{-1}$ \\
6.1 & $1.511 \times 10^{-1}$ & $0.711 \times 10^{-1}$ & $2.479 \times 10^{-1}$ \\
6.3 & $3.614 \times 10^{-1}$ & $0.928 \times 10^{-1}$ & $1.480 \times 10^{-1}$ \\
6.5 & $2.715 \times 10^{-1}$ & $0.644 \times 10^{-1}$ & $0.874 \times 10^{-1}$ \\
6.7 & $2.118 \times 10^{-1}$ & $0.350 \times 10^{-1}$ & $0.606 \times 10^{-1}$ \\
6.9 & $1.479 \times 10^{-1}$ & $0.213 \times 10^{-1}$ & $0.409 \times 10^{-1}$ \\
7.1 & $1.549 \times 10^{-1}$ & $0.159 \times 10^{-1}$ & $0.320 \times 10^{-1}$ \\
7.3 & $1.151 \times 10^{-1}$ & $0.104 \times 10^{-1}$ & $0.247 \times 10^{-1}$ \\
7.5 & $8.992 \times 10^{-2}$ & $0.724 \times 10^{-2}$ & $1.881 \times 10^{-2}$ \\
7.7 & $6.645 \times 10^{-2}$ & $0.420 \times 10^{-2}$ & $1.169 \times 10^{-2}$ \\
7.9 & $5.805 \times 10^{-2}$ & $0.359 \times 10^{-2}$ & $0.836 \times 10^{-2}$ \\
8.1 & $4.226 \times 10^{-2}$ & $0.171 \times 10^{-2}$ & $0.703 \times 10^{-2}$ \\
8.3 & $3.880 \times 10^{-2}$ & $0.123 \times 10^{-2}$ & $0.568 \times 10^{-2}$ \\
8.5 & $3.249 \times 10^{-2}$ & $0.084 \times 10^{-2}$ & $0.324 \times 10^{-2}$ \\
8.7 & $3.033 \times 10^{-2}$ & $0.075 \times 10^{-2}$ & $0.212 \times 10^{-2}$ \\
8.9 & $2.688 \times 10^{-2}$ & $0.045 \times 10^{-2}$ & $0.160 \times 10^{-2}$ \\
9.1 & $2.324 \times 10^{-2}$ & $0.032 \times 10^{-2}$ & $0.129 \times 10^{-2}$ \\
9.3 & $1.855 \times 10^{-2}$ & $0.022 \times 10^{-2}$ & $0.095 \times 10^{-2}$ \\
9.5 & $1.339 \times 10^{-2}$ & $0.015 \times 10^{-2}$ & $0.075 \times 10^{-2}$ \\
9.7 & $8.957 \times 10^{-3}$ & $0.098 \times 10^{-3}$ & $0.604 \times 10^{-3}$ \\
9.9 & $4.994 \times 10^{-3}$ & $0.056 \times 10^{-3}$ & $0.453 \times 10^{-3}$ \\
10.1 & $2.190 \times 10^{-3}$ & $0.029 \times 10^{-3}$ & $0.256 \times 10^{-3}$ \\
10.3 & $7.037 \times 10^{-4}$ & $0.114 \times 10^{-4}$ & $1.083 \times 10^{-4}$ \\
10.5 & $1.560 \times 10^{-4}$ & $0.032 \times 10^{-4}$ & $0.316 \times 10^{-4}$ \\
10.7 & $2.168 \times 10^{-5}$ & $0.133 \times 10^{-5}$ & $0.557 \times 10^{-5}$ \\
10.9 & $2.982 \times 10^{-6}$ & $0.440 \times 10^{-6}$ & $0.839 \times 10^{-6}$ \\
\hline
\end{tabular}}
\begin{flushleft}
\textbf{Notes.} The measured HIMF and its best-fitting Schechter function are shown in Figure\,\ref{Fig:himf}, with both the statistical and systematic errors displayed.
\end{flushleft}
\end{table}

\subsection{Cosmic \hi density and systematic uncertainty}
\label{sec:omegaHI_systematics}

\begin{figure*}
\centering
\includegraphics[width=0.49\textwidth, angle=0]{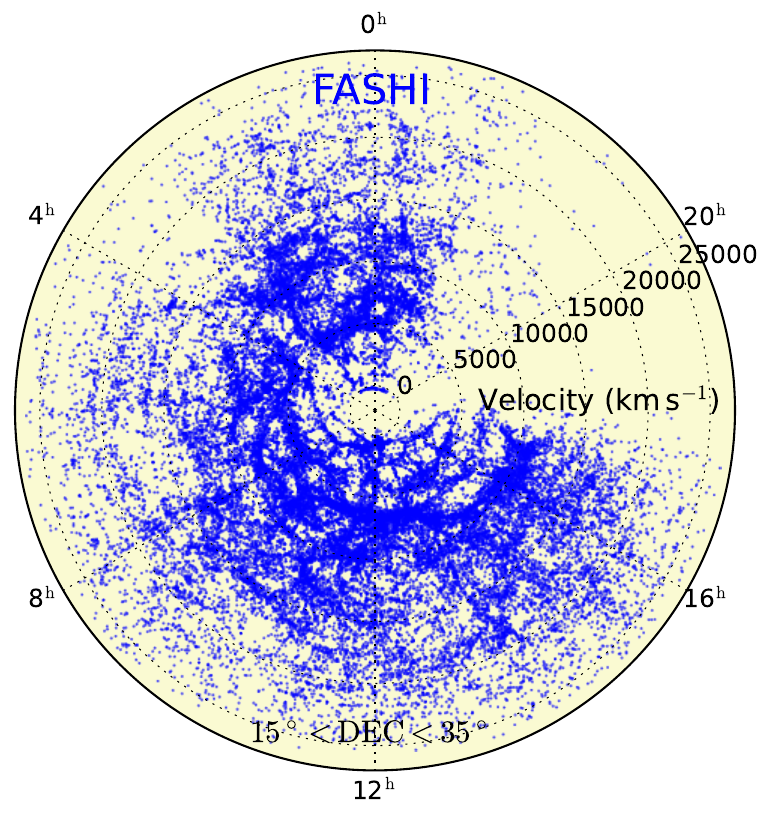}
\includegraphics[width=0.49\textwidth, angle=0]{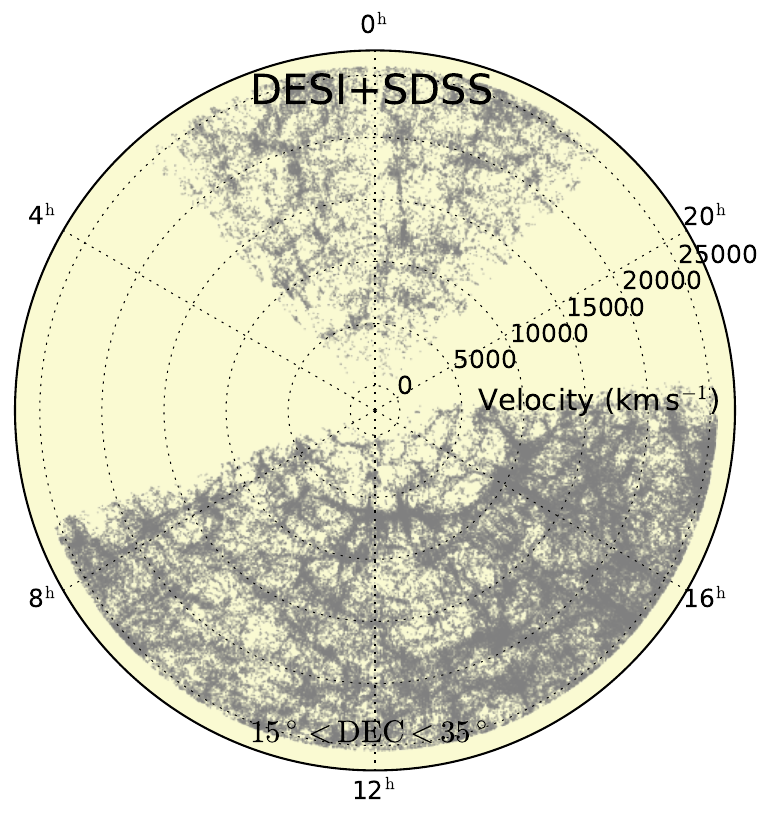}
\includegraphics[width=0.49\textwidth, angle=0]{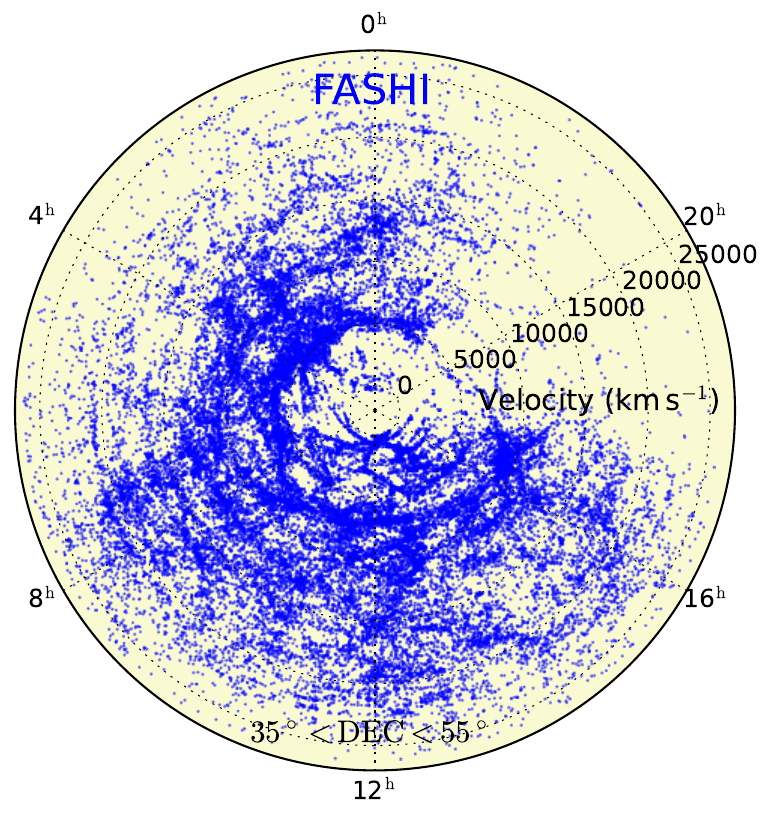}
\includegraphics[width=0.49\textwidth, angle=0]{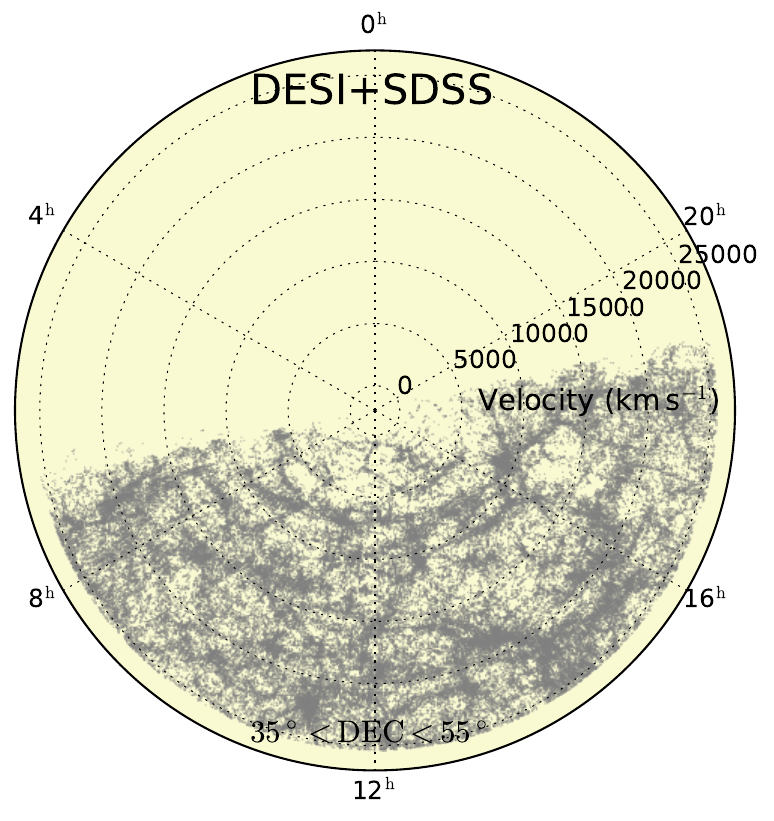}
\caption{Cosmic web traced by radio (FASHI \hi sources) and optical (DESI and SDSS galaxies) surveys spanning two declination strips: $\rm 15^\circ < \text{DEC} < 35^\circ$ and $\rm 35^\circ < \text{DEC} < 55^\circ$ (J2000). The sample is volume-limited to $z = 0.09$, a cutoff corresponding to the maximum velocity in the current FASHI release.}
\label{Fig:polar}
\end{figure*}

The large-scale structure of the Universe is well traced by the spatial distribution of galaxies observed in extensive optical spectroscopic surveys such as DESI and SDSS, which map the cosmic web in three dimensions well beyond the local Universe \citep[e.g.,][]{Abazajian2009,desi2025}. Within the FASHI survey footprint at $z<0.09$, we detect approximately 156\,000 \hi galaxy sample. As shown in Figure\,\ref{Fig:polar}, their spatial distribution broadly follows the filamentary and cellular patterns delineated by optically selected galaxy populations from DESI and SDSS, including prominent filaments, walls, and voids.

\begin{table*}
\centering
\small
\caption{\bf{Systematic uncertainty budget for $\Omega_{\rm HI}$.}}
\label{tab:omegaHI_sys}
\vskip 5pt
\begin{tabular}{lcc} 
\hline\hline
Source &
$\sigma(\Omega_{\rm HI})\,(10^{-4}\,h_{70}^{-1})$ &
Fractional uncertainty \\
\hline
Statistical uncertainty & 0.03 & 0.6\% \\
\hline
Flux measurement uncertainty & 0.002 & 0.05\% \\
Absolute flux calibration & 0.09 & 1.9\% \\
Distance uncertainty & 0.005 & 0.11\% \\
50\% completeness limit & 0.34 & 7.2\% \\
Joint measurement uncertainty & 0.37 & 7.9\% \\
Jackknife sample variance & 0.15 & 3.2\% \\
Bias-corrected cosmic variance & 0.79 & 16.8\% \\
\hline
Primary systematic: joint MC $+$ jackknife & 0.40 & 8.5\% \\
Conservative systematic: joint MC $+$ CV & 0.87 & 18.5\% \\
\hline
\end{tabular}
\begin{flushleft}
\footnotesize
\textbf{Notes.}
The fractional uncertainties are calculated relative to the nominal value
$\Omega_{\rm HI}=4.71\times10^{-4}\,h_{70}^{-1}$. The primary systematic uncertainty combines the joint Monte Carlo measurement uncertainty with the empirical jackknife sample variance. The bias-corrected conservative estimate replaces the jackknife term with the Driver-Robotham cosmic-variance estimate after rescaling by the relative bias of H\,{\sc i}-selected and optically selected galaxies.
\end{flushleft}
\end{table*}

We further estimate the cosmic \hi density by integrating the measured HIMF:
\begin{equation}
    \Omega_\hj = \frac{1}{\rho_c} \int M_\hj \, \phi(M_\hj) \, dM_\hj,
\end{equation}
where $\rho_c$ is the critical density at $z=0$. Given the broad mass range covered by our sample, we perform a direct summation over the observed HIMF and obtain $\Omega_\hj = (4.71 \pm 0.03_{\rm stat}) \times 10^{-4}\,h_{70}^{-1}$, where $0.03_{\rm stat}$ denotes the statistical error. The statistical error in FASHI DR2 is much smaller than that of DR1 \citep{Ma2025}, due to the significantly larger sample. Galaxies with $M_\hj < 10^7\,\msun$ contribute only 1.5\% of the total $\Omega_\hj$. Hence, the larger uncertainties at the low-mass end do not significantly affect the precision of the cosmic \hi density measurement. In contrast, galaxies with $M_\hj > 10^9\,\msun$ account for more than 75\% of the total $\Omega_\hj$.

With the significantly reduced statistical uncertainty of the HIMF measurement, the systematic uncertainty estimate becomes much more important. Following \cite{Jones2018}, we consider the systematic uncertainties from flux measurement, flux calibration, luminosity distance measurement, 50\% completeness limit $S_{\hj,50\%}$, sample variance and cosmic variance in $\Omega_\hj$. 

We first estimate the measurement-related systematic uncertainty using a joint Monte Carlo approach. In each Monte Carlo realization, we simultaneously perturb the integrated \hi fluxes, luminosity distances, the absolute flux calibration, and the parameters of $S_{\hj, 50\%}(W_{50},f_\sigma)$ relation in Eq.~(\ref{eq:shi50}). The details are as follows.

The perturbed \hi flux is written as $S_{\hj,i}^\prime=(S_{\hj,i}+\delta S_i)(1+\delta_{\rm cal})$, where $\delta S_i$ is drawn from the flux uncertainty of each source and flux calibration uncertainty $\delta_{\rm cal}$ is drawn from a Gaussian distribution with a width of 5\%. The luminosity distance of each source is perturbed according to its quoted uncertainty, $D_i^\prime = D_i + \delta D_i$, with $\delta D_i$ being the Gaussian scatter.

The uncertainty in the $S_{\hj, 50\%}(W_{50},f_\sigma)$ comes from the uncertainties in the seven best-fitting parameters of Eq.~(\ref{eq:shi50}). It is propagated using the covariance matrices of the fitted parameters. We draw parameters of Eq.~(\ref{eq:shi50}) from their multivariate Gaussian distributions. This preserves the covariance and avoids treating the fitted parameters as independent nuisance parameters. For each realization, we recompute $M_{\rm HI}$, $S_{\hj, 50\%}$, $V_{\max}$, the HIMF, and $\Omega_{\rm HI}$. The scatter of the resulting $\Omega_{\rm HI}$ distribution gives
\begin{equation}
\sigma_{\rm meas}(\Omega_{\rm HI})=0.37 \times 10^{-4}h_{70}^{-1}.
\end{equation}

Another systematic effect is the sample variance, i.e., the variation of HIMF within the survey volume, which can be estimated directly from the FASHI footprint using jackknife resampling. The survey area ($\sim$$19\,500\deg^2$) is divided into 100 sky regions of equal area, and the HIMF and $\Omega_{\rm HI}$ are recalculated after removing each region in turn \citep{Jones2018}. The resulting jackknife uncertainty is
\begin{equation}
\sigma_{\rm JK}(\Omega_{\rm HI})=0.15 \times 10^{-4}h_{70}^{-1},
\end{equation}
which is much smaller than the measurement uncertainty $\sigma_{\rm meas}$ owing to the much larger volume of FASHI DR2.

We adopt our primary estimate of the systematic uncertainty $\sigma_{\rm sys}$ by combining the joint Monte Carlo measurement uncertainty and the empirical jackknife sample variance in quadrature,
\begin{equation}
\sigma_{\rm sys}=\left(\sigma_{\rm meas}^2+\sigma_{\rm JK}^2\right)^{1/2}=0.40 \times 10^{-4}h_{70}^{-1}.
\end{equation}

Besides all the above systematic uncertainties, we further estimate the possible contribution from cosmic variance of finite survey volume, due to the fluctuations in the large-scale density field \citep{Moster2011}. We adopt the empirical prescription of \citet{Driver2010} (their Eq.~4), which was calibrated using optically selected galaxies in SDSS with $\pm1$~mag of $M_r-5\log h_{70}=-21.58$~mag. Their original formula gives the fractional cosmic variance for a survey volume specified by its three effective dimensions and its aspect ratio. However, FASHI is a flux-limited \hi survey, so different \hi masses are sampled over different effective volumes. We therefore do not apply the formula of \cite{Driver2010} to a single survey volume. Instead, we evaluate a mass-dependent, source-weighted cosmic-variance contribution using the same accessible volume that enters the HIMF calculation.

The FASHI footprint is divided into connected angular patches using the pixelized sky mask. For each galaxy in a patch, we evaluate the fractional cosmic variance using its own radial depth and  the transverse dimensions of the patch. Each source, and hence each \hi mass range, is assigned a cosmic-variance amplitude appropriate to its own flux-limited accessible volume.

The cosmic-variance contribution is then propagated using the same weights as in the HIMF and $\Omega_\hj$ estimators. For the HIMF, the contribution of each source to the cosmic variance is $w_i=\sigma_{{\rm CV},i}/V_{\max,i}$. For $\Omega_\hj$, the weight becomes $w_i=\sigma_{{\rm CV},i}M_{\hj,i}/V_{\max,i}$. We assume that the large-scale density fluctuation is coherent within each connected sky patch but independent between different patches. The contributions to the cosmic variance of each patch are added in quadrature to derive the final cosmic variance for HIMF and $\Omega_\hj$. This treatment explicitly accounts for the fact that low-mass \hi galaxies are detected over smaller effective radial depths, while massive \hi galaxies probe a larger fraction of the survey volume. It also avoids treating individual galaxies within the same large-scale structure as independent realizations. The resulting estimate of cosmic variance gives $\sigma_{\rm CV,DR10}=1.35 \times 10^{-4}h_{70}^{-1}$.

The prescription was calibrated for optically selected, approximately $M_\star$-like galaxies. Since \hi-selected galaxies are more weakly clustered, the uncorrected estimate is expected to overestimate the cosmic variance of an \hi-selected sample. At fixed survey volume and window function, the fractional cosmic-variance amplitude scales approximately linearly with the tracer bias \citep{Moster2011},
\begin{equation}
\sigma_{{\rm CV},\hj}=\frac{b_\hj}{b_{\rm opt}}\sigma_{\rm CV,DR10}.
\end{equation}
The characteristic luminosity of the calibration in \cite{Driver2010} corresponds to approximately $M_r-5\log h\simeq -20.8$, for which SDSS clustering measurements imply $b_{\rm opt}\simeq1.2$ \citep[see e.g., Fig.~7 of][]{Zehavi2011}. Observational measurements of \hi-selected galaxies give $b_{\hj}\simeq0.7$ \citep[e.g.][]{Li2025}. With this correction, the cosmic-variance contribution to $\Omega_\hj$ is $\sigma_{{\rm CV},\hj}=0.79\times10^{-4}h_{70}^{-1}$.

Combining this bias-corrected cosmic-variance estimate with the joint Monte Carlo measurement uncertainty gives the conservative systematic uncertainty
\begin{equation}
\sigma_{\rm sys,CV}=\left(\sigma_{\rm meas}^2+\sigma_{\rm CV,HI}^2\right)^{1/2}=0.87\times10^{-4}h_{70}^{-1}.
\end{equation}
We quote this value as a conservative estimate including cosmic variance. Our primary systematic uncertainty, however, is based on the joint Monte Carlo measurement uncertainty combined with the empirical jackknife sample variance measured directly from the FASHI footprint.

The systematic uncertainty budget is summarized in Table~\ref{tab:omegaHI_sys}. The first four components are also evaluated separately as diagnostics, but they are not added independently in quadrature to define the fiducial measurement uncertainty. Instead, the joint Monte Carlo scatter is used as the measurement-related systematic uncertainty.

With the primary systematic uncertainty, our final measurement is
\begin{equation}
\Omega_{\rm HI}=\left(4.71\pm 0.03_{\rm stat}\pm 0.40_{\rm sys}\right)\times10^{-4}h_{70}^{-1}.
\end{equation}
The effective redshift is calculated as, 
\begin{equation}
\langle z\rangle=\sum_i z_iM_{\hj,i}\phi_i(M_{\hj})/\sum_iM_{\hj,i}\phi_i(M_{\hj}),
\end{equation}
which gives $\langle z\rangle=0.027$ for the FASHI DR2 sample.
Including the bias-corrected cosmic-variance estimate gives a conservative systematic uncertainty of
\begin{equation}
\sigma_{\rm sys,CV}=0.87\times10^{-4}h_{70}^{-1}.
\end{equation}
Compared to the systematic uncertainties estimated for ALFALFA in \cite{Jones2018}, we further include the uncertainties from the 50\% completeness limit which dominates the measurement errors. In \cite{Jones2018}, the error from the absolute flux calibration is quoted as the systematic uncertainty for $\Omega_\hj$, at the level of $0.6\times10^{-4}h_{70}^{-1}$. With the improved sensitivity and much larger volume of the FASHI survey, the systematic uncertainties in the HIMF and $\Omega_\hj$ are substantially reduced.

\begin{figure}[H]
\centering
\includegraphics[width=0.48\textwidth]{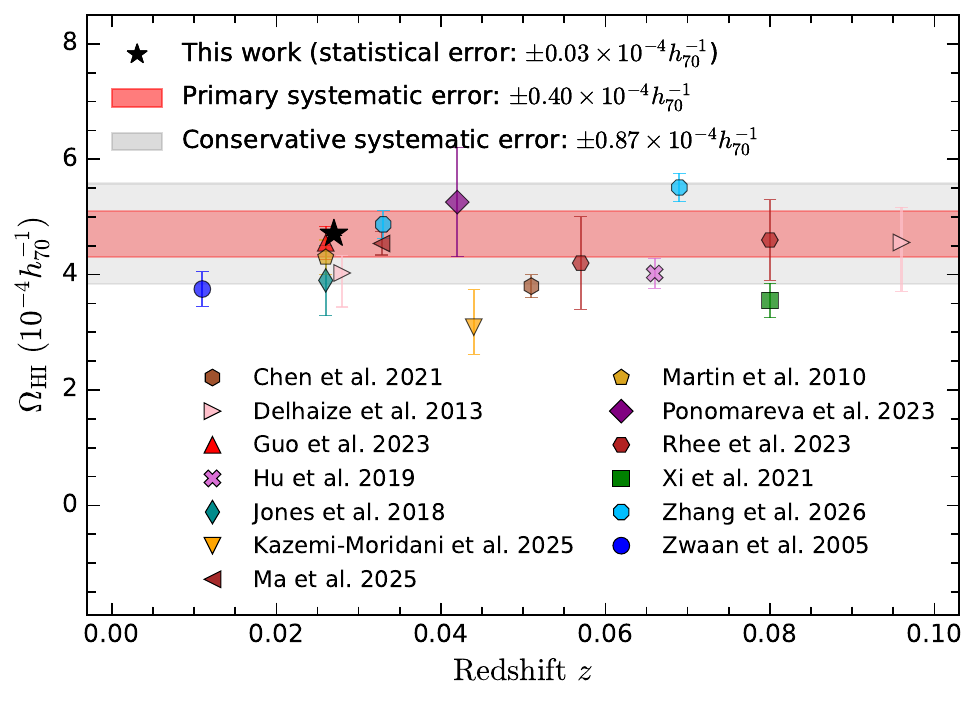}
\caption{Evolution of the cosmic \hi\ gas density ($\Omega_\hj$) with redshift. Our measurements are shown as a black filled star, with statistical and systematic uncertainties indicated. For comparison, we overplot previous 21-cm observational results from the literature \citep{Zwaan2005,Xi2021,Guo2023,Ponomareva2023,Ma2025,Delhaize2013,Hu2019,Chen2021,Rhee2023,ZhangNA2026}. All measurements have been homogenized to a common set of cosmological parameters.}
\label{Fig:omega}
\end{figure}

In Figure~\ref{Fig:omega}, we show the comparisons of our $\Omega_\hj$ measurement with the previous measurements derived from HIMFs \citep{Zwaan2005,Martin2010,Jones2018,Guo2023,Ponomareva2023,Kazemi-Moridani2025,Ma2025} and \hi-stacking measurements \citep{Delhaize2013,Hu2019,Chen2021,Xi2021,Rhee2023}. Our primary systematic uncertainty is shown as the red band, while the conservative systematic uncertainty including the cosmic variance is shown as the gray band. Our $\Omega_\hj$ is in reasonable agreement with these previous measurements over the redshift range of $0<z<0.1$. The evolution of $\Omega_\hj$ tends to be weak in this redshift range. However, we note that only a few measurements include systematic uncertainties in the quoted errors \citep{Zwaan2005,Jones2018}.

\subsection{Large-scale Structure}

The large-scale structure of the Universe is well traced by the spatial distribution of galaxies observed in extensive optical spectroscopic surveys such as DESI and SDSS, which map the cosmic web in three dimensions well beyond the local Universe \citep[e.g.,][]{Abazajian2009,desi2025}. Within the FASHI survey footprint at $z<0.09$, we detect approximately 156\,000 \hi galaxy sample. As shown in Figure\,\ref{Fig:polar}, their spatial distribution broadly follows the filamentary and cellular patterns delineated by optically selected galaxy populations from DESI and SDSS, including prominent filaments, walls, and voids.

However, a more detailed comparison reveals notable differences between the \hi-traced and optically traced cosmic web. The Sloan Great Wall (SGW) at $z\sim0.08$, a prominent structure in optical surveys, is not prominently detected in FASHI, partly due to the sparse sampling of \hi sources at these distances. As shown in \cite{Einasto2011}, majority of the SGW galaxies are in the red sequence, with the $r$-band absolute magnitude of $M_r<-20.5$ and $g-r$ color of $g-r>0.75$. These galaxies corresponds to massive galaxies of $M_\star\sim10^{10.7}\msun$, which are expected to have a median \hi mass around $10^{9.2}\,\msun$ \citep{Catinella2018}. The expected flux density of these galaxies at $z\sim0.08$ is only $0.05\,{\rm Jy}\,\kms$, which is below the FASHI detection limit. Only the most \hi-rich SGW galaxies at these redshifts can be observed. Conversely, the local voids ($0<z<0.03$) identified in optical surveys \citep[e.g.,][]{Moster2011,Chen2019} are not apparent in the FASHI sample. This is consistent with previous findings that \hi-selected galaxies preferentially reside in underdense regions \citep{Guo2017}.

\begin{figure}[H]
\centering
\includegraphics[width=0.48\textwidth]{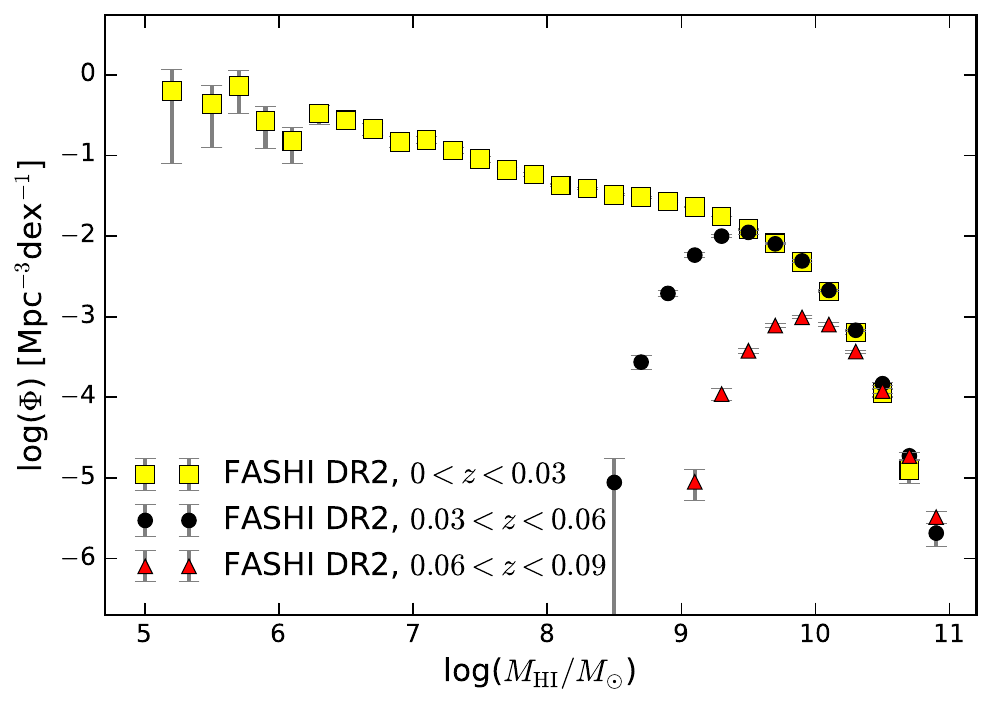}
\caption{H\,{\sc i} mass functions (HIMFs) in three redshift bins: $0<z<0.03$, $0.03<z<0.06$, and $0.06<z<0.09$.}
\label{Fig:himf_z}
\end{figure}

To further quantify this, we present the HIMFs in three redshift bins ($0<z<0.03$, $0.03<z<0.06$, and $0.06<z<0.09$) in Figure\,\ref{Fig:himf_z}. In mass ranges where the statistics are complete ($M_\hj>10^{9.5}\,\msun$ for $0.03<z<0.06$ and $M_\hj>10^{10.5}\,\msun$ for $0.06<z<0.09$), we observe no decrease in number density toward the lowest redshift bin. This indicates that the optical local voids do not correspond to a deficit of \hi galaxies, further supporting the environmental dependence of \hi content. More detailed studies of the influence of the cosmic web on \hi distribution will be presented in future work.

As seen from the average 50\% mass completeness limits in Figure~\ref{Fig:mass_distance}, for galaxies above $M_\hj=10^{10}\msun$, majority of the galaxies are complete over the whole redshift range of $0<z<0.09$.
Figure\,\ref{Fig:himf_z} shows that the HIMF exhibits very little evolution at the massive end over $0<z<0.09$, suggesting a balance between \hi accretion and depletion over this redshift range. This finding is consistent with previous measurements of the HIMF at $z\gtrsim0.05$ based on much smaller samples \citep{Xi2021,Ponomareva2023}. As the FASHI survey continues to improve in sensitivity and completeness, future analyses will enable quantitative studies of the \hi cosmic web over significantly larger volumes.

\subsection{Caveats and Comments}

Despite the generally high quality of the FASHI data products, several caveats should be considered when using the source catalog presented in Table~\ref{tab:exgalcat}. These issues primarily affect a small fraction of sources or specific derived parameters, but they are summarized here for completeness and to guide appropriate scientific usage.

\begin{figure}[H]
\centering
\includegraphics[width=0.48\textwidth, angle=0]{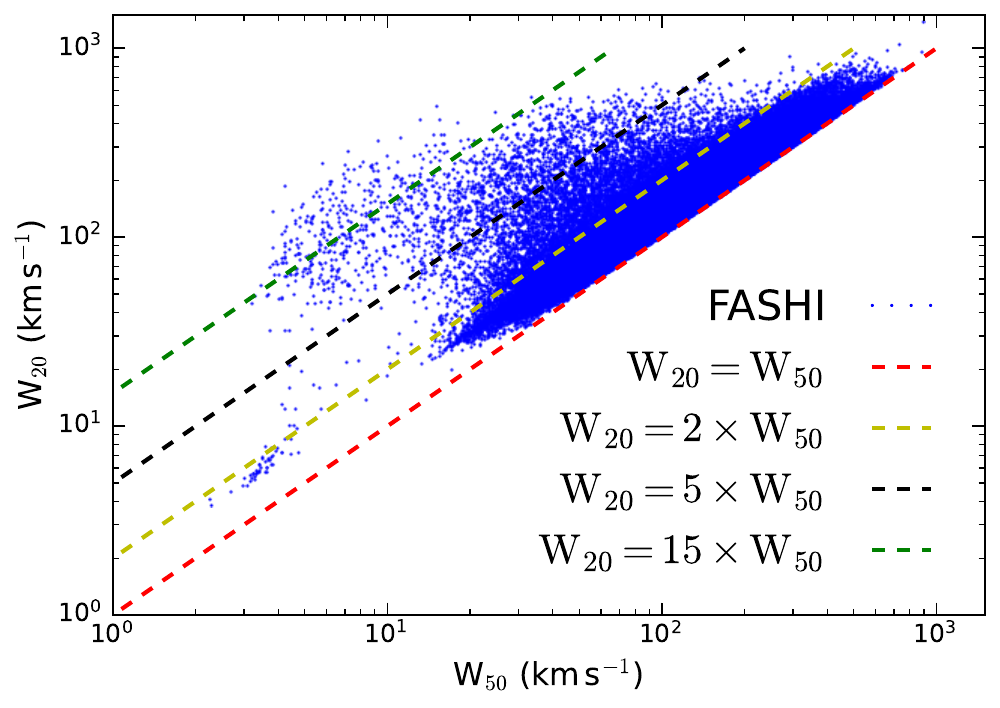}
\caption{Comparison of $W_{20}$ and $W_{50}$ line widths measured via busy-function fitting. Here, $W_{20}$ and $W_{50}$ are defined at 20\% and 50\% of the peak flux density, respectively. Points with a high width ratio ($W_{20}/W_{50} > 5$) correspond to double-peaked profiles with a strong intensity contrast. Points with both widths below $10\,\kms$ originate from single-peaked profiles probably affected by RFI near the spectral peak.}
\label{Fig:w20_w50}
\end{figure}

First, for a subset of double-peaked \hi spectra, the line width measured at 50\% of the peak flux density ($W_{50}$), derived using the busy-function fitting, can be significantly smaller than the corresponding $W_{20}$ when the secondary peak has an intensity below 50\% of the primary peak (see Figure\,\ref{Fig:w20_w50}). In such cases, $W_{50}$ may not accurately reflect the full velocity extent of the \hi profile. Users are therefore advised to exercise caution when employing $W_{50}$ for kinematic analyses or statistical studies, and to apply additional corrections or alternative width definitions where appropriate. $1569$ galaxies ($1.01\%$) in our sample have $W_{20}>5W_{50}$, and for the massive end of $\log(M_\hj/\msun)>10$, only 121 galaxies ($0.63\%$ of the massive galaxies) are above this cut. Therefore, it only has a minor effect on our HIMF measurement.

Second, approximately 500 candidates initially identified by the source finder were found to have low reliability, owing to low signal-to-noise ratios (${\rm SNR}<4$) or proximity to the edges of the velocity (spectral channel) or spatial (image pixel) coverage. These low-reliability sources have been excluded from the released FASHI catalog. Follow-up optical or \hi observations will be required to assess their nature and improve their characterization.

Third, the current public release focuses exclusively on \hi emission-line sources. A small number of negative features consistent with \hi absorption have been identified in the data cubes \citep[e.g.,][]{Zhang2025}, but these are not included in the present catalog. Dedicated data reduction and validation procedures for absorption systems are ongoing and will be presented in future work.

Fourth, because the FASHI survey footprint includes regions at low Galactic latitude, residual contamination from Galactic radio recombination lines (RRLs) may persist despite careful flagging and inspection \citep[e.g.,][]{Zhang2021,Hou2022}. Furthermore, extragalactic OH megamasers (OHMs) have been excluded from the catalog \citep[e.g.,][]{Zhang2024oh} only when a matching spectroscopic redshift is available in DESI or SDSS; unidentified OHMs without optical redshift information may therefore remain in the sample. To quantify this potential contamination, we examined the $\sim$64\,000 FASHI sources with DESI/SDSS spectroscopic counterparts and identified 18 sources with spectral characteristics consistent with OHMs, representing $\sim$0.028\% of the spectroscopically covered sample. Extrapolating to the $\sim$92\,000 sources without spectroscopic coverage yields an expected $\sim$26 undetected OHMs (only $\sim$0.017\% of the full catalog). Users conducting studies sensitive to such contaminants should apply additional screening based on spectral characteristics or cross-matching with ancillary catalogs.

Fifth, the DR2 catalog incorporates nearly all sources reported in FASHI DR1 \citep{Zhang2024}, with the exception of $\sim$1\,500 objects that are not recovered in the updated observations. Given the higher sensitivity and improved data quality of DR2, these unrecovered sources are most likely spurious detections in DR1, arising from positional uncertainties in extended sources or from residual RFI contamination. Such suspicious entries have been removed from the DR2 catalog through updated processing and validation procedures. Based on our validation process combining multi-criteria inspection, the false-detection rate for the full DR2 catalog is estimated to be below 2\%. We therefore recommend that users preferentially adopt the DR2 catalog for scientific analyses. For applications requiring the highest reliability, users may select sources with higher SNR or secure optical spectroscopic counterparts.

Sixth, the spatial distribution of \hi flux within individual sources is often irregular for the extended sources, and can also be affected by residual radio-frequency interference or nearby objects. As a result, the centroid positions derived from the \hi data may deviate from the true optical positions by more than the nominal FAST pointing accuracy of $\sim$15$''$ \citep{Jiang2020}. In practice, however, the positional offsets are typically smaller than $1'$ as demonstrated by the separations between FASHI detections and their DESI or SDSS counterparts. While this level of uncertainty is negligible for most statistical studies, it should be taken into account in detailed analyses of individual systems or in high-precision cross-matching applications.

Seventh, in the current DR2 catalog, sources detected by the blind SoFiA pipeline (Section\,\ref{sec:sofia}) and those identified through optically guided extraction (Section\,\ref{sec:opt_guide}) were merged prior to the final visual validation step. Consequently, the detection origin of individual sources is not preserved in the released catalog. Users requiring separate subsamples for their analyses should be aware of this limitation. For future data releases (DR3 and beyond), we will modify the processing workflow to preserve the extraction metadata and include a dedicated detection-method flag (with values blind/guided/both) in the released catalog.

Finally, RFI. The FASHI survey benefits from the regulatory protection of the FAST site. Satellite systems such as Starlink have not obtained operating licenses within China's radio regulatory framework and therefore do not provide active service coverage in this region; consistent with this, monthly monitoring of the electromagnetic environment over the 2020-2025 period shows no statistically significant increase in satellite-related RFI\footnote{\url{https://fast.bao.ac.cn/cms/category/rfi_monitoring_en/}}. Terrestrial RFI is effectively controlled by the strictly enforced 30\,km electromagnetic quiet zone around FAST. RFI levels are generally lower at night and higher during the day, and show a mild increase toward the celestial equator (DEC $\lesssim 3^\circ$ or DEC $\gtrsim 55^\circ$), likely due to geostationary satellites and low-elevation observations. A detailed frequency-dependent analysis is presented by \citet{Zhang2022}. As a fill-time survey, FASHI has non-uniform sky coverage; regions affected by elevated RFI are typically observed multiple times (in some cases $>$10 passes), ensuring that clean integrations can be retained. Our completeness model (Section~\ref{sec:completeness}), parameterised by per-source sensitivity $f_\sigma$ and line width $W_{50}$, naturally incorporates spatial variations in effective sensitivity, thereby capturing the dominant impact of RFI on source detectability. Consequently, the impact of RFI on the HIMF and $\Omega_{\rm HI}$ measurements is expected to be negligible.

\subsection{Accessing FASHI Data}
\label{sec:spectra}

In addition to the catalog listed in Table~\ref{tab:exgalcat}, integrated one-dimensional \hi line spectra for all cataloged sources have been extracted from the reduced three-dimensional FITS data cubes. The full FASHI \hi source catalog will be made publicly available upon publication of this paper.

The reduced one-dimensional spectral cubelets associated with individual sources are available upon request from the corresponding authors. Access to the reduced three-dimensional data cubes is currently restricted to internal collaboration use; however, a limited subset of these data may be provided upon reasonable request for specific scientific purposes.

All raw observational data from the FASHI project will be released to the public following a twelve-month proprietary period, in accordance with the FAST data release policy\footnote{The twelve-month proprietary period applies to each individual scan observation, rather than to the completion of the full survey.}. Further details regarding FAST data access and release policies are available at \url{https://fast.bao.ac.cn/cms/article/129/}.

\section{Summary}\label{sec:summ}

The FAST All Sky \hi Survey (FASHI) is a pioneering neutral hydrogen survey conducted using the world's most sensitive single-dish radio telescope, the Five-hundred-meter Aperture Spherical radio Telescope (FAST). Aimed at mapping the entire visible sky from FAST, covering declinations from $-14^\circ$ to $+66^\circ$ ($\sim$22\,000\,deg$^2$), FASHI observed approximately $19\,500\ \text{deg}^2$ ($47.3\%$ of the full sky) between August 2020 and July 2025. It has detected more than $156\,000$ extragalactic \hi sources at redshifts $z < 0.09$, with a median sensitivity of $0.57\ \text{mJy}\ \text{beam}^{-1}$ and a velocity resolution of $6.4\ \text{km}\ \text{s}^{-1}$ --- making it significantly deeper and more extensive than previous surveys such as ALFALFA.

Key outcomes of the survey include:

\begin{itemize}
    \item The construction of the largest uniform catalog of extragalactic \hi sources to date, with a detection rate of $\sim$8.0 sources per square degree.
    \item Cross‑matching with DESI and SDSS provides spectroscopic redshifts for $\sim$64\,000 sources, while $\sim$92\,000 sources lack spectroscopic coverage.
    \item The survey is sensitive to rare populations such as low-surface-brightness dwarfs, nearly dark galaxies, and diffuse intergalactic gas.
    \item Multi-wavelength synergies with eROSITA and optical surveys allow studies of the baryon cycle, environmental effects, and the cosmic web.
    \item Using a completeness‑corrected sample of over \(109\,000\) galaxies, we derive the HIMF robustly down to $M_{\mathrm{HI}}\sim10^{6.2}\,M_\odot$. Measurements below this mass are shown for completeness but are affected by Local Volume structure and are not used to infer the global low-mass slope. Including systematic uncertainties, the HIMF is well described by a single-Schechter function, with no statistically robust requirement for a second Schechter component. We obtain a cosmic \hi density \(\Omega_{\mathrm{HI}}=(4.71\pm0.03_{\rm stat}\pm0.40_{\rm sys})\times10^{-4}h_{70}^{-1}\), with a conservative systematic uncertainty of $0.87\times10^{-4}h_{70}^{-1}$ when cosmic variance is included.
    \item FASHI \hi sources broadly trace the optical cosmic web, yet optical local voids are not devoid of \hi galaxies, confirming that \hj-selected systems preferentially reside in underdense environments. The HIMF shows no significant evolution at the massive end over \(0<z<0.09\).
\end{itemize}

The paper presents the survey design, observational strategy, data reduction pipeline (\texttt{HiFAST}), source extraction methodology, and catalog validation. Comparative analyses with ALFALFA demonstrate FASHI's superior sensitivity and reliability, especially for faint sources. The released catalog includes detailed source parameters such as positions, velocities, line widths, fluxes, distances, and \hi masses, and is publicly accessible for community use.

FASHI establishes a legacy dataset that bridges the gap between earlier shallow surveys and future SKA-era projects, providing critical insights into gas content, galaxy evolution, and large-scale structure in the local Universe.

\section*{Acknowledgements}

We thank the anonymous referee for their careful reading and constructive comments that helped improve the clarity and quality of this paper. This work is supported by the National Natural Science Foundation of China (12225303, 12288102), the National SKA Program of China (2025SKA0150100), the National Key R\&D Program of China (2025YFE0202300), the National Natural Science Foundation of China (12373001, 12595313, 12173045), the Science Research Grants from the China Manned Space Project (CMS-CSST-2021-A05, CMS-CSST-2025-A07), the Guizhou Provincial Science and Technology Projects (QKHFQ[2023]003, QKHPTRC-ZDSYS[2023]003, QKHFQ[2024]001, QKHJCMS[2025]015), the Chinese Academy of Sciences (CAS) Project for Young Scientists in Basic Research (YSBR-063, YSBR-092), and the CAS through a grant to the CAS South America Center for Astronomy, and the Chinese Academy of Sciences South America Center for Astronomy (CASSACA) Key Research Project (E52H540101, E52H540301). FAST is a Chinese national mega-science facility, operated by the National Astronomical Observatories of Chinese Academy of Sciences (NAOC).


\section*{Conflict of Interest}
The authors declare that they have no conflict of interest.


{\small \setlength{\baselineskip}{-1pt}
\bibliographystyle{raa}
\bibliography{references}
}

\end{multicols}

\end{document}